\documentclass[aps,prd,twocolumn,showpacs,showkeys,superscriptaddress,amsmath,amssymb]{revtex4}
\usepackage[usenames,dvipsnames]{color}
\usepackage{epsf}
\usepackage{bm}
\usepackage{dcolumn}
\usepackage{latexsym}
\usepackage{amsmath}
\usepackage{amsfonts}
\usepackage{amssymb}
\usepackage{graphicx}
\usepackage[active]{srcltx}
\newcommand{\be}{\begin{eqnarray}}
\newcommand{\ee}{\end{eqnarray}}

\newcommand{\bi}{\hat{b}}

\def\rmd{{\rm{d}}}
\def\rmi{{\rm{i}}}
\def\rme{{\rm{e}}}
\def\bm#1{\mbox{\boldmath{$#1$}}}
\def\bi#1{{\bm #1}}
\newcommand{\revision}{}
\definecolor{darkred}{rgb}{.8,0,0}

\definecolor{darkblue}{rgb}{0,0,.7}

\begin{document}
%
\title{The emergence of Special and Doubly Special Relativity}
%
%
\author{Petr Jizba}
\email{p.jizba@fjfi.cvut.cz}
\affiliation{FNSPE, Czech Technical
University in Prague, B\v{r}ehov\'{a} 7, 115 19 Praha 1, Czech Republic\\}
\affiliation{ITP, Freie Universit\"{a}t Berlin, Arnimallee 14
D-14195 Berlin, Germany}
\author{Fabio Scardigli}
\email{fabio@phys.ntu.edu.tw}
\affiliation{
Leung Center for Cosmology and Particle Astrophysics (LeCosPA),
Department of Physics, National Taiwan University, Taipei~106, Taiwan\\}
\affiliation{Yukawa Institute for Theoretical Physics, Kyoto
University, Kyoto 606-8502, Japan}

\begin{abstract}
\vspace{3mm}
 \begin{center}
{\bf Abstract}\\[2mm]
\end{center}
Building on our previous work [Phys.~Rev.~D{\bf 82},~085016~(2010)], we show in this paper how a
Brownian motion on a short scale can originate
a relativistic motion on scales that are larger than particle's Compton wavelength. 
This can be described in terms of polycrystalline vacuum.
Viewed in this way,
special relativity is not a primitive concept, but rather it statistically emerges when a coarse
graining average over distances of order, or longer than the Compton wavelength is taken.
By analyzing the robustness of such a special relativity under small variations in the polycrystalline 
grain-size distribution we naturally arrive at the notion of doubly-special
relativistic dynamics. In this way, a previously unsuspected, common statistical origin of the two
frameworks is brought to light. Salient issues such as the r\^{o}le of gauge fixing in emergent relativity, 
generalized commutation relations, Hausdorff dimensions of representative path-integral
trajectories and a connection with Feynman chessboard model are also discussed.
%
%
%
%

%
\end{abstract}
%
\maketitle
%
%
\section{Introduction~\label{sec1}}
Without doubts the discovery of
Lorentz symmetry (LS) irrevocably changed the theoretical landscape
in physics. Up to now LS has been confirmed to unprecedent
precision, and during the last century it has powerfully constrained
theories in a way that has proved instrumental in discovering new
laws of physics. Moreover, the mathematical structure of the Lorentz
group is compellingly simple and elegant. It thus seems natural to assume that
Lorentz invariance is an exact symmetry of nature which is valid for
an arbitrary boost. Yet, there are several reasons to doubt the
exactness of LS. From a purely conceptual standpoint, the most
cogent reason is that an infinite volume of the Lorentz group is
experimentally untestable since, unlike the
rotation group, the Lorentz group is non-compact. Why should one
then assume that exact LS holds when this hypothesis cannot be
tested, not even in principle? The non-compactness
may, indeed, sound as a logically convincing argument for doubting the exactness of LS but
in itself is not enough to attract a sufficient attention.
There are, however, other more pressing reasons to suspect that LS
may fail at some critical energy or boost. For instance, in  quantum
field theory both the ultraviolet divergences and Landau poles are
direct artifacts of the assumption that the spectrum of field
degrees of freedom is boost invariant. Another sound reason comes
from quantum gravity where profound difficulties associated with the
problem of time~\cite{Isham,Isham II,Kuchar} indicate that an
underlying preferred time may be necessary
in order to reconcile gravitational and quantum physics. In
particular, general arguments imply that a radical departure from
standard space-time symmetries at the Planck scale~\cite{thooft-a} is
necessary. Aside from general issues of principle, specific hints of
Lorentz violation come from tentative calculations in various
approaches to quantum gravity. Examples include: space-time
foam~\cite{hawking}, cosmologically varying moduli~\cite{Damour}, Witten's
string field theory~\cite{Kostelecky II}, semiclassical spin-network calculations in
Loop quantum gravity~\cite{Gambini,Alfaro}, non-commutative
geometry~\cite{Hayakawa,Carrol,Anisimov}, world-crystal
physics~\cite{HKII,JKS1}, 't~Hooft's cosmic cellular automata~\cite{thooft}
or condensed matter analogues of emergent gravity~\cite{Volovik03}.
None of the above reasons amount to a convincing argument that a LS breaking is an
inevitable aspect of quantum gravity. However, taken together they do motivate serious
attempts to address possible observable consequences of a violation
of LS, and to strengthen observational bounds. What should be perhaps
emphasized is that the idea of LS violation is not
new and it has been considered by a number of authors over the last
forty years or so (see, e.g. Refs.~\cite{fock, DSR} and citations therein).
It has, however, received a serious boost only during the past decade. The
catalyst has been both a massive infusion of ideas from quantum gravity,
and improvements in observational sensitivity that allow to detect violations
of LS that are linearly Planck suppressed (see e.g.,~\cite{jacobson:08} for an
extensive review).

In this paper we show that a relativistic
quantum mechanics, as formulated through path integrals (PI), bears
in itself a seed of understanding how LS can be broken at short
spatio-temporal scales and yet emerge as an apparently exact
symmetry at large scales. Our argument is based upon a recent
observation~\cite{JK1,JK2} that PI for both fermionic and bosonic
relativistic particles may be interpreted (when analytically continued
to imaginary times) as describing a doubly-stochastic process that operates on two vastly
different spatio-temporal scales.
The short spatial scale, which is much smaller than
the Compton length, describes a Wiener (i.e., non-relativistic)
process with a fluctuating Newtonian mass. This might be visualized as if the particle would
be randomly propagating (in the sense of Brownian motion) through a granular or ``polycrystalline" medium.
The large spatial scale corresponds, on the other hand, to distances that are much larger
than particle's Compton length. At such a scale the particle evolves according to a genuine
relativistic motion, with a sharp value of the mass coinciding with the Einstein rest mass.
Particularly striking is the fact that when we average the particle's velocity over the
correlation distance (i.e., over particle's Compton wavelength) we obtain the velocity of light $c$.
So the picture that emerges from this analysis is that the particle (with a non-zero mass!)
propagates over the correlation distance $~ 1/mc$ (hereafter $\hbar = 1$)
with an average velocity $c$, while at larger
distance scales (i.e., when a more coarse grained view is taken) the particle propagates as
a relativistic particle with a sharp mass and an average velocity that is smaller than $c$.
This bears a strong resemblance with Feynman's chessboard PI for a relativistic Dirac fermion
in $1+1$ dimensions~\cite{FH}. There, an analogous situation occurs, i.e., a {\em massive}
particle propagates over distances of Compton length with velocity $c$, and it is only on
much larger spatial scales where the Brownian motion with a sub-luminal average
velocity emerges~\cite{FH,Schulman-Jacobson}.  The analogy with Feynman's chessboard PI appears
also on the level of Hausdorff dimensions of representative trajectories. While below the
Compton wavelength the Haussdorff dimension $d_H=1$, which corresponds to a super-diffusive
process, on scales much larger than the Compton length one has $d_H=2$, which is the usual Brownian
diffusion.
In passing, we may stress that the outlined superposition of two stochastic
processes with widely separated times scales fits the conceptual framework which is often
referred to as a superstatistics~\cite{Beck:01}.

Of course, a single-particle relativistic quantum
theory is a logically untenable concept, since a multi-particle production is allowed whenever
the particle reaches the threshold energy for pair production. At the same time, the PI for a
single relativistic particle is a perfectly legitimate building block in quantum field
theory (QFT). Indeed, QFT can be viewed as a grand-canonical ensemble of particle histories
where Feynman diagrammatic representation of quantum fields depicts directly the pictures
of the world-lines in a grand-canonical ensemble. In particular, the partition function for
quantized relativistic fields can be fully rephrased in terms of single-particle
relativistic PI's. This view is epitomized, e.g., in the Bern--Kosower ``string-inspired"
approach to quantum field theory~\cite{Bern-Kosower}
or in Kleinert's disorder field theory~\cite{KleinertIII}.

The outlined scenario can be also conveniently applied in various doubly special relativistic
(DSR) models. In those models a further invariant scale $\ell$, besides the speed of light $c$,
is introduced, and $\ell$ is assumed typically to be of the order of the Planck length.
In the present framework, the scale $\ell$ can be naturally identified with the minimal grain
size of the polycrystalline medium.
By following the same strategy as in the special relativistic context, i.e., analyzing the structure of paths
which enter the Feynman summation, one can again identify correlation lengths, canonical
commutation relations and the respective Hausdorff dimensions. All of these critically
depend on the DSR model at hand, and may serve to gain insight into the underlying stochastic
process which is, as a rule, related by an analytic continuation with the corresponding
quantum mechanical dynamics.

The purpose of this paper is to call attention to such a peculiar behavior of relativistic PI
at short spatio-temporal scales --- a fact already recognized by Feynman --- and bring ensuing
implications to the attention of our particle-physics and cosmology colleagues.

The structure of the paper is as follows. To set the stage we recall in the next section some
fundamentals of Markovian smearing of path integrals, also known as  {\em superstatistics}
path integrals (SPI). Section~\ref{sec3} is devoted to application of SPI in relativistic
quantum mechanics. We restrict our presentation largely to bosons of zero spin that are described
by the Klein--Gordon equation. Thought the Klein--Gordon particle (KGP) is not a key for the
results obtained, it will allow to elucidate the physics behind our reasonings quite
straightforwardly. In particular we show how a transitional amplitude for the KGP can be
written as a superposition of non-relativistic free-particle PI's with different Newtonian
masses. To this end we use a less known but equivalent representation of Klein--Gordon
equation, namely the so-called Feshbach--Villars representation. The concept of emergent
relativity is discussed in Section~\ref{sec4}. There we observe that the superstatistics
version of Feynman path summation for a relativistic particle allows the following
probabilistic interpretation: the single-particle
relativistic theory might be viewed as a single-particle {\em non\/}-relativistic theory
(Wiener process) whose Newtonian mass $\tilde{m}$ (which is not invariant under Lorentz
transformations) is a fluctuating parameter, whose average approaches the true relativistic
Einstein mass $m$ at observational (or resolution) times that are much larger than the Compton time $1/mc^2$. On a spatial
scale greater than the particle's Compton wave length the particle follows the standard
relativistic motion with a sharp mass and a sub-luminal average velocity.
%
%
Sections~\ref{Sec.Vabc} and \ref{sec6ab} contain discussions of two conceptually important concomitant topics.
In particular, Section~\ref{Sec.Vabc} concentrate on the issue of stability of the emergent special relativity (SR) under a small perturbation of the grain (or mass-smearing) distribution in the polycrystalline vacuum.
There we show that small perturbations naturally lead to the DSR theory. Hence the class of DSR models (of which the special relativity theory is a particular example) is robust under
and a small change of the grain-size distribution. In Section~\ref{sec6ab}, we are concerned with the question
of how a choice of a gauge fixing influences our polycrystalline-vacuum picture. We employ the
St\"{u}ckelberg field-enlarging trick and show that the SPI in question can be made explicitly
invariant under gauge transformations (or reparametrizations), i.e., under the same group under which
relativistic particle systems are invariant.
By revealing that our original action is dynamically equivalent to the relativistic action with the reparametrization symmetry
we are allowed to proclaim the ``polycrystalline" picture as being a basic (or primitive) edifice of SR, and consider the reparametrization symmetry as a mere artifact of an artificial redundancy that is allowed in the description.
%
%


In Section~\ref{sec5}, we extend our approach to doubly special relativistic dynamics,
sketch the computation of the ensuing canonical commutation relations (CCR's) and the
Hausdorff dimensions of representative trajectories. Since the smearing distribution
implicitly corresponds to the gauge fixing condition, the obtained CCR automatically
match the quantized Dirac brackets. It is, indeed, a bonus of the superstatistics
PI for (doubly-)relativistic particle that it directly provides a symplectic structure
in the reduced phase space. We close Section~\ref{sec5} with some comments on the
underlying non-relativistic picture.
%

Various remarks and generalizations are proposed in the concluding section.
For the reader's convenience the paper is supplemented with four appendices which clarify
some finer technical details.

\section{Superstatistics path integrals\label{sec2}}

We begin with the
well known fact that when a conditional probability density
functions (PDF's) is formulated through PI then it satisfies
the Chapman--Kolmogorov equation (CKE) for continuous Markovian
processes, namely
\begin{eqnarray}
 P(x_b, t_b|x_a,t_a)=  \int_{-\infty}^{\infty} \rmd x\ \!
 P(x_b,
t_b|x, t) P(x, t |x_a,t_a)\, .
\label{II1a}
\end{eqnarray}
Conversely, any probability satisfying CKE possesses a PI
representation~\cite{FH,Kac}.

In physics one often encounters probabilities formulated as a
superposition of PI's, e.g.
\begin{eqnarray}
&&\mbox{\hspace{-5mm}}\bar{P}(x_b,t_b|x_a,t_a)\nonumber \\
&&\mbox{\hspace{-5mm}}= \! \int_{0}^{\infty} \rmd v \  \omega(v,t_{ba})
\int_{x(t_a) = x_a}^{x(t_b) = x_b}{\mathcal{D}} x {\mathcal{D}} p\
e^{\int_{t_a}^{t_b} \rmd \tau \left(\footnotesize{\rmi} p\dot{x} - v
H(p,x) \right)}\, .\nonumber \\
\label{2a}
\end{eqnarray}
Here $\omega(v,t_{ba})$ with $t_{ba}=t_b-t_a$ is a normalized PDF
defined on ${\mathbb{{R}}}^+\!\!\times {\mathbb{{R}}}^+$. The random
variable $v$ is in practice typically related to the inverse temperature,
coupling constant, friction constant or volatility.
%

At this stage one may ask if is it possible that also $\bar{P}({\bi x}_b,t_b|{\bi x}_a,t_a)$
satisfies the CKE (\ref{II1a}). The answer is surprisingly affirmative provided $\omega (v,t)$
fulfills a certain simple functional equation. Following Ref.~\cite{JK1} we define a rescaled weight function
\begin{equation}
w(v,t)\ \equiv\  \omega (v/t,t)/t\, ,
\end{equation}
and calculate its
Laplace transform
\begin{eqnarray}
\tilde w(p_v,t)\ \equiv \
\int_0^{\infty} \rmd v  \ \! e^{-p_vv}w(v,t)\, .
\label{II4a}
\end{eqnarray}
Then $\bar{P}({\bi x}_b,t_b|{\bi x}_a,t_a)$ satisfies CKE only if
\begin{eqnarray}
\tilde w(p_v,t_1+t_2) \ = \ \tilde w(p_v,t_2) \tilde w(p_v,t_1)\, .
\end{eqnarray}
\vspace{2mm}
Assuming continuity in  $t$, $w(p_v,t)$ is unique and can be explicitly
written as (see~\cite{JK1}):
\begin{equation}
\tilde w(p_v,t)\ = \ [G(p_v)]^t \ = \
 e^{-t F(p_v)}\, .
 \label{II6a}
 \end{equation}
A function $F(p_v)$ must increase monotonically in order to allow for
inverse Laplace transform, and satisfy the  condition
$F(0)=0$ to ensure that $\omega$ is normalized to one.
Finally the Laplace inverse of
$\tilde w(p_v,t)$ yields $\omega(v,t)$.

Once the above conditions are satisfied, then  $\bar{P}({\bi x}_b,t_b|{\bi x}_a,t_a)$
possesses a path integral representation on its own.
The new Hamiltonian is given  by the relation
$\bar{{H}}({\bi p},{\bi x}) = F(H({\bi p},{\bi
x}))$. Here one must worry about the notorious
operator-ordering problem, not knowing in which temporal
order ${\bi p}$ and ${\bi x}$ must be taken in $F$.
At this stage it suffices to observe that when $H$ is ${\bi x}$ independent, the
former relation is exact. The issue of general $H$'s and the ensuing operator ordering
was discussed in detail in Ref.~\cite{JK1}.

\section{Path integral for Feshbach--Villars particle\label{sec3}}

Our following argument is based upon
Refs.~\cite{JK1,JK2} where it
%
was shown that the Newton--Wigner~\cite{NW,HK} propagator
for a relativistic scalar particle with Hamiltonian
$\hat{H}({\bi p}) = c\sqrt{\hat{\bi p}^2 + m^2c^2}$, i.e.
\begin{widetext}
\begin{eqnarray}
P({\bi x}_b,t_b|{\bi x}_a,t_a)\ = \ \int_{\footnotesize{\bi x}(t_a)\ \! =
\ \! \footnotesize{\bi x}_a}^{\footnotesize{\bi
x}(t_b) \ \! = \ \! \footnotesize{\bi x}_b}
{\mathcal{D}}{\bi x} \frac{\mathcal{D} {\bi p}}{(2\pi)^D} \ \!
\exp\left\{\int_{t_a}^{t_b}\!\! \rmd \tau \ \!\left[\rmi {\bi p}
\cdot\dot{\bi x} - c\sqrt{{\bi p}^2 + m^2 c^2}\right]\right\}\, ,
\label{8a}
\end{eqnarray}
%
can be considered as a superposition of  non-relativistic
free-particle path integrals provided one chooses the generating function
$G(p_v) \ = \ e^{-a\sqrt{p_v}}$ with  $a \in \mathbb{{R}}^+$.
In such a case one obtains~\cite{JK1,JK2}
%
%
\begin{eqnarray}
P({\bi x}_b,t_b|{\bi x}_a,t_a)
\ = \ \int_{0}^{\infty}\!\!\!\rmd v \
\! \omega(v,t_{ba}) \int_{\footnotesize{\bi x}(t_a)\ \! = \ \! \footnotesize{\bi x}_a}^{\footnotesize{\bi
x}(t_b) \ \! = \ \! \footnotesize{\bi x}_b}
{\mathcal{D}}{\bi x}  \  \! \frac{\mathcal{D} {\bi
p}}{(2\pi)^D} \ \! \exp\left\{\int_{t_a}^{t_b}\!\!\! \rmd \tau\
\![\rmi {\bi p} \cdot\dot{\bi x} - v ({\bi p}^2 c^2 + m^2
c^4)]\right\}\, ,
\label{9a}
\end{eqnarray}
\end{widetext}
with $\omega(v,t)$ being the Weibull distribution of order $1$. In
general, Weibull's PDF of order $a$ is defined as~\cite{weibull}
\begin{eqnarray}
\omega(v,a,t) \ = \ \frac{a \exp\left(-a^2t/4v\right)}{2\sqrt{\pi}
\sqrt{v^3/t}}\, .
\label{IIIbaa}
\end{eqnarray}
We note that, neither (\ref{8a}) nor (\ref{9a}) are propagators for
Klein--Gordon equation. As stressed first by
Stuckelberg~\cite{Stuckelberg:41,Stuckelberg:42}, the true
relativistic propagator must include also the negative energy spectrum,
reflecting the existence of charge-conjugated solutions, i.e.,
antiparticles.

Recently, it was pointed out that the resolution of this problem in
the framework of PI's can be readily found when the Klein--Gordon particle
is written in the so-called Feshbach--Villars (FV)
representation~\cite{JK2}
\begin{eqnarray}
&&i\partial_t \Psi \ = \ \hat{H}_{_{\rm FV}} ({\bi p}) \Psi\, ,\nonumber \\
&&\hat{H}_{_{\rm FV}} ({\bi p}) \ = \ (\sigma_3  +  \rmi\sigma_2)
\frac{\hat{\bi p}^2}{2m} \ + \ \sigma_3mc^2\, ,
\label{III10a}
\end{eqnarray}
where $\Psi$ is a two component wave function. The two components are related to
opposite parity states --- fact that is automatically fulfilled by Dirac bispinors
in case of Dirac's equations.
FV representation was already thoroughly discussed in Ref.~\cite{JK2} and we shall
refrain from going to further details here. The interested reader is referred to
Appendix~A where the relevant essentials are presented. {\revision
Here we only mention that in order to deal with the full PI representation of the
Klein--Gordon particle it will suffice to discuss the PI relation (\ref{9a}) alone,
see Eq.~(\ref{15a}).
}

\section{Emergent Special Relativity\label{sec4}}

If we now consider the
change of variable $vc^2 \leftrightarrow 1/2\tilde{m}$, then the RHS
of the key relativistic PI identity (\ref{9a}) can be rewritten in
the form [see also Ref.~\cite{JK2}]
\begin{widetext}
\begin{eqnarray}
 &&\mbox{\hspace{-15mm}}\int_{\footnotesize{\bi x}(0)\ \! =
 \ \! \footnotesize{\bi x}'}^{\footnotesize{\bi
x}(t) \ \! = \ \! \footnotesize{\bi x}} \ \!{\mathcal{D}}{\bi x} \frac{\mathcal{D}
{\bi p}}{(2\pi)^D} \ \! \exp\left\{\int_{0}^{t} \!\!\rmd \tau \
\!\left[\rmi {\bi p}\cdot \dot{\bi x} \ - \
c\sqrt{{\bi p}^2 + m^2 c^2}\right]\right\} \nonumber \\[2mm]
&&\mbox{\hspace{15mm}}= \  \int_{0}^{\infty}\!\!\rmd \tilde{m} \
\sqrt{\frac{c^2 t}{2\pi\tilde{m}}} \ \ \rme^{-tc^2(\tilde{m} - m)^2/2\tilde{m}}
\int_{\footnotesize{\bi x}(0)\ \! = \ \! \footnotesize{\bi x}'}^{\footnotesize{\bi
x}(t) \ \! = \ \! \footnotesize{\bi x}} \ \!{\mathcal{D}}{\bi x} \ \! \frac{\mathcal{D} {\bi
p}}{(2\pi)^D} \ \! \exp\left\{\int_{0}^{t} \!\!\rmd \tau\
\!\left[\rmi {\bi p}\cdot \dot{\bi x} \ - \  \frac{{\bi p}^2}{2
\tilde{m}} \ - \  m c^2 \right]\right\} \nonumber \\[2mm]
&&\mbox{\hspace{15mm}}= \  \int_{0}^{\infty}\!\!\rmd \tilde{m} \
\! f_{\frac{1}{2}}\!\left(\tilde{m}, tc^2, tc^2m^2\right)
\int_{\footnotesize{\bi x}(0)\ \! = \ \! \footnotesize{\bi x}'}^{\footnotesize{\bi
x}(t) \ \! = \ \! \footnotesize{\bi x}} \ \!{\mathcal{D}}{\bi x} \ \! \frac{\mathcal{D} {\bi
p}}{(2\pi)^D} \ \! \exp\left\{\int_{0}^{t} \!\!\rmd \tau\
\!\left[\rmi {\bi p}\cdot \dot{\bi x} \ - \  \frac{{\bi p}^2}{2
\tilde{m}} \ - \  m c^2 \right]\right\}\, ,
\label{22a}
\end{eqnarray}
\end{widetext}
where $t_b-t_a \equiv t-0 = t$, and
\begin{eqnarray}
f_p(z,a,b) \ = \ \frac{(a/b)^{p/2}}{2K_p(\sqrt{a b})} \ \! z^{p-1} \
\!\rme^{-(az + b/z)/2}\, ,
\label{23a}
\end{eqnarray}
is the generalized inverse Gaussian distribution~\cite{feller66}
($K_p$ is the modified Bessel function of the second kind with index
$p$). The structure of (\ref{22a}) suggests that $\tilde{m}$ can be
interpreted as a Newtonian mass which takes on continuous values
distributed according to $f_{\frac{1}{2}}\!\left(\tilde{m}, tc^2,
tc^2m^2\right)$ with $\langle \tilde{m} \rangle = m + 1/tc^2$ and
$\mbox{var}(\tilde{m}) = m/tc^2 + 2/t^2c^4$. As a result one may
view a single-particle relativistic theory as a single-particle
non-relativistic theory where the particle's Newtonian mass $\tilde{m}$
represents a fluctuating parameter which approaches on average the
Einstein rest mass $m$ in the large $t$ limit.
We stress that the time $t$ in question should be understood
as a time after which the observation
(i.e., the position measurement) is made. In particular, during the period $t$ the system remains unperturbed.
In this respect the smearing distribution
$f_{\frac{1}{2}}\!\left(\tilde{m}, tc^2, tc^2m^2\right)$ represents a
temporal coarse-grained distribution for a Newtonian mass ---
the longer the time between measurements, the poorer the resolution of mass fluctuations.
One can thus justly expect that in the long run all mass fluctuations will be
washed out and only a sharp time-independent effective mass will be perceived.
The form of $\langle \tilde{m} \rangle$ identifies
the time scale at which this happens with $ t \sim 1/mc^2$. The latter is the
time for light to cross the particle's Compton wavelength --- i.e.,
the Compton time $t_C$. The expression for
$\langle \tilde{m} \rangle$ suggests however also another
interesting physical implication.
As we have seen, when  $t \gg  1/mc^2$  then
$\langle \tilde{m} \rangle$ rapidly converges to the
relativistic value $m$, signaling that the motion becomes genuinely
relativistic at large times. Note that for $t$ large enough we surely
have $m > 1/tc^2$ which we can read as $mc^2 t > 1$.
The latter means that, for large $t$, the relativistic Heisenberg
inequality for the energy/time variables is satisfied, $\Delta E
\Delta t \geq 1$. On the other hand, for $t \ll 1/mc^2$, the fluctuations of the Newtonian
mass $\tilde{m}$ around the average $m$ are huge. The motion takes place inside a
specific space-grain, and in each space-grain the motion is
a classical, i.e. non relativistic, Brownian motion controlled by the Hamiltonian
${\bi p}^2 / 2 \tilde{m}$. There the relativistic Heisenberg
uncertainty relation is clearly violated, in fact $mc^2 t < 1$
(remind that $m$ is the Einstein rest mass).
However, if we compute the non-relativistic Heisenberg relation, using the Newtonian
mass $\tilde{m}$ and the non-relativistic kinetic energy $E_{kin} \sim \tilde{m}v^2$, we find
\begin{eqnarray}
\Delta t  \Delta E_{kin} \ = \ \left\langle \tilde{m} \frac{(\Delta x)^2}{\Delta t} \right\rangle \ \sim \ 1\, .
\label{IV13a}
\end{eqnarray}
So the non-relativistic Heisenberg relation is not violated.
This is because for a Brownian motion
the standard non-relativistic scaling $(\Delta x)^2/\Delta t \sim 1/\tilde{m}$ holds.
In this connection it is interesting to observe that for $t \ll 1/mc^2$ we have
$\langle \tilde{m} \rangle \sim 1/(\Delta t~ c^2 )$. By comparing this with (\ref{IV13a})
we see that on a short time scale
$\Delta x \sim c \Delta t$. So the corresponding stochastic process is
super-diffusive. The preceding scaling behavior will be
rigorously justified in Appendices~B and C.

The observant reader might notice that the PI (\ref{22a}) can be identified with a PI for a relativistic
particle in, the so-called, Polyakov's gauge~\cite{polyakov:87}.
So, the form of the  smearing function $f_{\frac{1}{2}}\!\left(\tilde{m}, tc^2, tc^2m^2\right)$
naturally fixes the gauge, which in this case turns out to be to Polyakov's gauge.
%
In this way the use of
smearing functions bypasses the conventional Dirac--Bergman methodology for quantization
of constrained systems. In fact, in Appendix~B
we arrive at the correct special-relativistic CCR
\begin{eqnarray}
[\hat{\bi x}_j, \hat{\bi p}_i] \ = \
\rmi\left(\delta_{ij} + \frac{\hat{\bi p}_i \hat{\bi p}_j}{m^2 c^2}\right)\, ,
\end{eqnarray}
without using the machinery of Dirac brackets.

Fluctuations of the Newtonian mass can be depicted as originating
from particle's evolution in an ``inhomogeneous'' or a
``polycrystalline''  medium. Granularity, as well known, for instance, from
solid-state systems, typically leads to corrections in the local
dispersion relation~\cite{Johnson:93} and hence to alterations in
the local effective mass. The following picture thus emerges: on the
short-distance scale, a non-relativistic particle can be
envisaged as propagating through a single grain with a local mass
$\tilde{m}$, in a classical Brownian motion.
This fast-time process
has a time scale $\sim 1/{\tilde{m}}c^2$. An averaged
value of the time scale can be computed with the help of the smearing distribution
$f_{\frac{1}{2}}\!\left(\tilde{m}, tc^2, tc^2m^2\right)$,
which gives a transient temporal scale $\langle1/\tilde{m}c^2\rangle = 1/mc^2$.
The latter coincides with particle's
Compton time $t_C$.
At time scales much longer than $t_C$ (large-distance scale),
the probability that the particle encounters
a grain which endows it with a mass $\tilde{m}$ is
$f_{\frac{1}{2}}\!\left(\tilde{m}, tc^2, tc^2m^2\right)$. Because
the fast-time scale motion is essentially Brownian, the local
probability density matrix (PDM) conditioned on some fixed
$\tilde{m}$ in a given grain is Gaussian
\begin{eqnarray}
\hat{\rho}({{\bi p}},t|\tilde{m}) \ =
\  \left(\frac{t}{\tilde{m} 2\pi}\right)^{\!\!3/2}\ \! \exp\left[- t\frac{{\hat{\bi
p}}^2}{2\tilde{m}}\right] .
\end{eqnarray}
As the particle moves through a ``grainy environment" the Newtonian
mass $\tilde{m}$ fluctuates
and the corresponding joint PDM will be $\hat{\rho}({\bi
p},t;\tilde{m}) = f_{\frac{1}{2}}\!\left(\tilde{m}, tc^2,
tc^2m^2\right)\hat{\rho}({\bi p},t|\tilde{m})$.
%
%
The marginal
PDM describing the mass-averaged (i.e. long-term) behavior is thus
\begin{eqnarray}
\hat{\rho}({\bi p},t) = \int_{0}^{\infty} \rmd \tilde{m} \ \!
f_{\frac{1}{2}}\!\left(\tilde{m}, tc^2,
tc^2m^2\right)\hat{\rho}({\bi p},t|\tilde{m})\, .
\label{25a}
\end{eqnarray}
The matrix elements of $\hat{\rho}({\bi p},t)$ in the ${\bi x}$-basis are then described by
the PI (\ref{22a}).

\begin{figure}[ht]
    \begin{center}
    \includegraphics[width=0.5\textwidth]{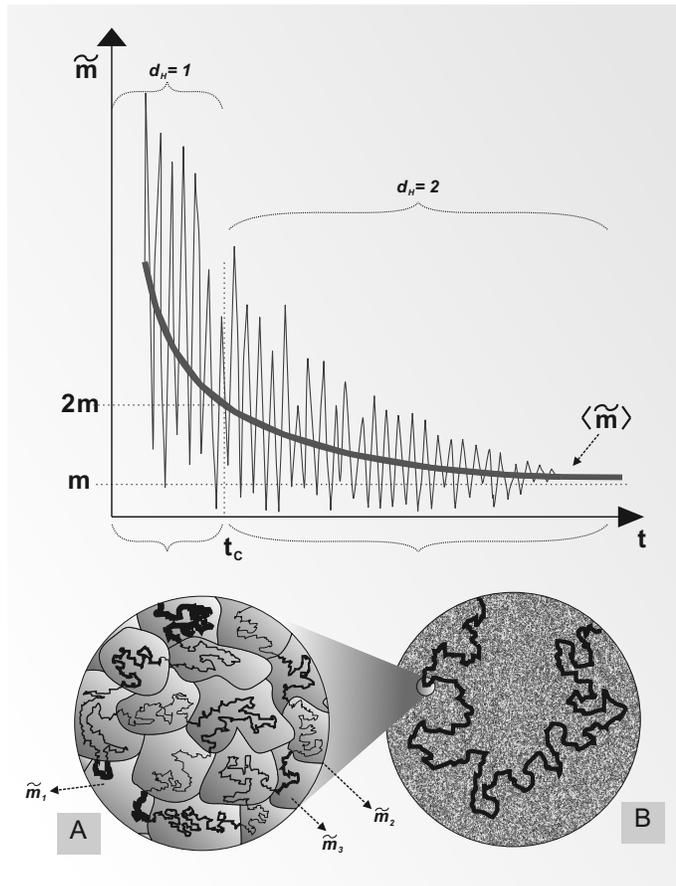}
    {\small \caption{The roughness of the representative trajectories in the relativistic
    path integral (\ref{22a}) depends on a
    spatial/temporal scale. On a fine scale (A), where
$t \ll t_C$ (or $\ell \ll \lambda_C$) a particle can be considered as propagating with a sharp
Newtonian mass $\tilde{m}$ in a single spatial grain with a Brownian motion controlled
by the Hamiltonian ${\bi p}^2/2\tilde{m}$.
On the intermediate scale of order $\lambda_C$ the particle propagates with an average velocity
equal to the speed of light $c$.
On a coarser scale (B)
the particle appears to follow a Brownian process with a sharp Lorentz invariant mass $m$, and the particle's
net velocity is then less than $c$ --- as it should be for a massive relativistic particle
governed by the relativistic Hamiltonian $c({\bi p}^2 + m^2 c^2)^{1/2}$.}
    \label{fig1}}
    \end{center}
\end{figure}

We may also observe that the averaged (or coarse-grained) velocity over the
correlation time $t = 1/mc^2$ equals the speed of light $c$. In fact
\begin{eqnarray}
\mbox{\hspace{-9mm}}&&\langle |{\bi v}| \rangle_{t = 1/mc^2}   \ = \
\left.\frac{\langle |{\bi p}| \rangle}{\langle \tilde{m} \rangle} \right|_{t = 1/mc^2}\nonumber \\[1mm]
\mbox{\hspace{-9mm}}&&= \  \frac{1}{2m}\int_{0}^{\infty} \!\!\!\!\rmd \tilde{m} \ \!
f_{\frac{1}{2}}\!\left(\tilde{m}, 1/m,
m\right) \sqrt{\left(\frac{8 m \tilde{m}c^2}{\pi } \right)} \
= \ c\, .
\end{eqnarray}
So on a short-time scale of order $\lambda_C$ the Klein--Gordon particle
propagates with an averaged velocity which is the speed of light $c$. But if one
checks the particle's position at widely separated intervals (much larger than $\lambda_C$),
then many directional reversals along a typical PI trajectory will take place, and the particle's
net velocity will be then less than $c$ --- as it should be for a massive particle (see Fig.~\ref{fig1}).
In addition, the time-compounded smearing distribution tends for large times
rapidly to the delta-function distribution $\delta(\tilde{m} -m)$
thanks to the central limit theorem. This means that the particle acquires a sharp
mass equal to Einstein's (i.e., Lorentz invariant) mass, and the process (not being
hindered by fluctuating masses) turns out to be purely Brownian. As detailed in Appendix B, this is
also confirmed by a direct calculation of the Feynman--Hibbs scaling
relation between $\Delta x$ and $\Delta t$ which indeed gives the fractal dimension
$2$ --- one of the key signatures of a Brownian motion.
%
%
%
%
%
%


On more formal level, a stochastic process, representing a
relativistic motion on the long-time scale, and described by the
Kramers--Moyal equation with the Kramers--Moyal operator
$c\sqrt{\hat{\bi p}^2 + m^2 c^2}$, is shown to be equivalent to a
doubly stochastic process in which the fast-time dynamics of a free
non-relativistic particle (Brownian motion) is coupled with a
long-time dynamics describing fluctuations of the particle's
Newtonian mass~\cite{JK1}.

On more speculative vein, one can fit the above observation into the
currently much debated emergent relativity theory, i.e. the approach that
tries to view either special or general theory of relativity not as
primitive concepts but rather as theories that statistically emerge
from a deeper (essentially non-relativistic) level of
dynamics~\cite{bohm:96,bohm:93,douglas:01,jacobson:08,Laughlin03,Froggatt91,Bjorken01,Volovik03}.

All these remarks extend directly also
to certain interacting systems. For instance,
to Dirac's Hamiltonian~\cite{JK2}
\begin{eqnarray}
{H}_{\rm D}^{A,V} =  c\gamma_0 \bm{\gamma} \cdot ({\bi p} -  e
{\bi A}/c)  +  \gamma_0 (m c^2 + V)  +  e A_0\, ,\;\label{iv.18}
\end{eqnarray}
and to the Feshbach--Villars Hamiltonian~\cite{JK2}
\begin{eqnarray}
&&\mbox{\hspace{-10mm}}{H}_{\rm FV}^{A,V}  =  (\sigma_3 + \rmi \sigma_2)\frac{1}{2m}({\bi
p} - e {\bi A}/c)^2\nonumber \\[2mm]  &&\mbox{\hspace{10mm}}+  \sigma_3 (m c^2 + V) + e A_0\, .
\label{iv.19}
\end{eqnarray}
For example, in the case when $V=0$, $A_x = -B y$ ($B_z \equiv B$), and $A_y = A_z = 0$,
then the PI for Dirac's Hamiltonian
yields the ``fast scale" Hamiltonian [see~\cite{JK2} for details]
\begin{eqnarray}
H_{\rm SP}  = \frac{1}{2 \tilde{m}}\left[ \left(p_x + \frac{e}{c}
B y \right)^{\! 2}   +  p_y^2  +  p_z^2 \right]  -  \mu_{\rm
B} B \sigma_3\, .
\end{eqnarray}
This is the Schr\"{o}dinger--Pauli Hamiltonian with $\mu_{\rm B} =
e\hbar/2\tilde{m}$ representing the Bohr magneton.  The
corresponding grain distribution is again the inverse Gauss
distribution. Analogous reasonings can be carried on also for charged
spin-$0$ particles, such as, e.g, $\pi^{\pm}$ mesons.

At first sight it may seem rather surprising that a LS process may
emerge from a superposition of two non-relativistic stochastic processes. What
is perhaps even more surprising is that none of the involved
processes has a dynamical symmetry that would correspond to a
Lorentz subgroup or to some form of a deformed Lorentz group
[see also~\cite{Calcagni, Reuter}]. This behavior is, however, less
striking when one observes that in many
doubly stochastic systems the statistically emergent behavior has a structure vastly
different from those of the respective defining processes.
Hydrodynamic turbulence provides an example, where the emergent
{\em velocity increments} and their ensuing Kolmogorov scaling can be understood
as originating from two stochastic processes (energy dissipation and chaotic force)
operating on two vastly different time scales, despite the fact that none of the
processes exhibits any particular scaling structure~\cite{beckII}. Analogous
situations are also known from financial
markets, e.g., credit risk models or stochastic volatility models.

\section{Robustness of SR under small variations
of mass-smearing distribution \label{Sec.Vabc}}

Let us now turn to the question how robust is the
emergent special relativity with respect to
a slight change  in the mass-smearing distribution.
In particular, we are asking what is the relation between
$\delta \omega$ (or equivalently $\delta f_{1/2}$) and $\delta F$.
Such a connection can be easily read off from the relations
(\ref{II4a}) and (\ref{II6a}). Namely,
\begin{eqnarray}
e^{-t [F(s) + \delta F(s)]}
\ = \ \int_0^{\infty} \rmd v \ \! e^{-tvs} [\omega(v,t) + \delta\omega(v,t)]\, ,
\end{eqnarray}
which directly implies
\begin{eqnarray}
-t e^{-tF(s)}\delta F(s)\ = \ \int_0^{\infty} \rmd v \ \! e^{-tvs}
\delta\omega(v,t)\, .
\label{VI33a}
\end{eqnarray}
Because of properties of the Laplace transform,
the only solution of Eq.~(\ref{VI33a}) for $\delta F = 0$
is $\delta\omega=0$. From this we immediately see that the smearing distribution
yielding the left-hand side of (\ref{22a}) is unique insofar as the
relativistic Hamiltonian for both positive and negative frequencies
has the usual square root form and the dynamics within a ``grain" is purely
Brownian. So the form of the SR is inexorably connected with the specific
structure of the mass-smearing distribution. The aforementioned specificity
might be, similarly as in the case of explicit values of the constants of Nature, attributed
either to the particular form of initial conditions
or to some as yet unknown dynamical mechanism.

The question naturally arising in this connection is how much the SR Hamiltonian $F(H) = \bar{H}({\bi p}) = c\sqrt{{\bi p}^2 + m^2c^2}$
changes when small perturbations around the mass-smearing distribution
(\ref{23a}) are considered. To answer this question we
restrict ourselves to the transformations of the type
\begin{eqnarray}
\hat{t} \ = \ t, \;\;\;\;\;\; \hat{v} \ = \ \hat{v}(v,t)\, .
\label{VII.32a}
\end{eqnarray}
Since $\omega$ should transform in $v$ as scalar density we have
\begin{eqnarray}
\hat{\omega}(\hat{v},{t}) \ = \ \frac{\partial v}{\partial \hat{v}} \ \! \omega(v,t)\, ,
\label{VII.33a}
\end{eqnarray}
or infinitesimally
\begin{eqnarray}
\delta \omega \ = \ \hat{\omega}(v,t) \ - \ \omega(v,t)
\ = \ -\frac{\partial}{\partial v}\left(\omega(v,t) \delta v \right)\, .
\label{VII.34a}
\end{eqnarray}
This also ensures that $\hat{\omega}$  is correctly normalized to $1$. In addition, we
require that $\delta \omega(v=0,t) =0$ and $\delta v(v=0,t) =0$, so that
$\hat{\omega}(v=0,t) = 0$ and the end-point $v=0$ is fixed.
Inserting (\ref{VII.34a}) back to Eq.~(\ref{VI33a}) we obtain equivalently
\begin{eqnarray}
\frac{e^{-tF(s)}}{s} \ \! \delta F(s)\ = \ \int_0^{\infty} \rmd v \ \! e^{-tvs}
\omega(v,t) \delta v\, .
\label{VII.36a}
\end{eqnarray}
Because of the condition $\delta v(v=0,t) =0$ we might assume that $\delta v$ can be represented by
the series
\begin{eqnarray}
\delta v(v,t) \ = \ v^{\alpha} \sum_{n=0}^{\infty} \epsilon_{n}(t)\ \! v^{n}\, ,
\label{VII.36abc}
\end{eqnarray}
where $\alpha\leq 1$ is a positive constant and $\epsilon_{n}(t) \ll 1$ (in order to facilitate small variations in $v$).
Through Eq.~(\ref{VII.34a}) this also implies that
\begin{eqnarray}
\delta \omega(v,t) \ =&&- \ \!\omega(v,t)v^{\alpha}\left[\sum_{n=0}^{\infty}v^n \left(\frac{t}{4v^2} - \frac{3}{2 v}\right)\epsilon_{n}(t) \right. \nonumber \\
&&+ \ \left. \sum_{n=0}^{\infty}v^{n -1} (n + \alpha ) \epsilon_{n}(t) \right],
\label{VII.37abc}
\end{eqnarray}
which indeed satisfies the required condition  $\delta \omega(v=0,t) = 0$.

Note further, that (\ref{VII.36a}) can be with the help of (\ref{VII.36abc}) written as
%
%
\begin{eqnarray}
\delta F(s)
  \ &=& \   \sum_{n=0}^{\infty} \epsilon_{n}(t)\ \! \left(\frac{1}{2\sqrt{s}}\right)^{n+\alpha}
\!\!  K_{n+\alpha-1/2}(t\sqrt{s})\nonumber \\[2mm]
&\times& \  s \ \! e^{t\sqrt{s}}  \sqrt{\frac{2 t \sqrt{s}}{\pi}} \, .
\label{VII.37abcd}
\end{eqnarray}
%
To reveal more details about the previous expansion, let us look at the large-$s$ asymptotic expansion. From theory of modified Bessel functions of the second kind it is known that
\begin{eqnarray}
K_\nu(x)\ = \ \sqrt{\frac{\pi}{2}} \frac{e^{-x}}{\sqrt{x}} \left(1 + \frac{4\nu^2 -1}{8x} + \mathcal{O}(1/x^2)\right) \, ,
\end{eqnarray}
at $x \rightarrow \infty $ and hence from (\ref{VII.37abcd}) the leading order terms are
\begin{eqnarray}
\delta F(s) \! \  &=&  \! \ s \! \ \epsilon_{0}(t) \!\left(\frac{1}{2\sqrt{s}}\right)^{\!\alpha} \nonumber \\[2mm] &+& \! \ s  \left(\frac{1}{2\sqrt{s}}\right)^{\!\alpha + 1}\!\left[\epsilon_{1}(t) + \frac{\epsilon_{0}(t) (\alpha^2 -\alpha)}{t} \right]\nonumber \\[2mm]
&+&  \! \ \mathcal{O}(s^{-\alpha/2})\,  .
\label{V.31.ab}
\end{eqnarray}
So in order to have the RHS $t$ independent, the leading-order coefficient  $\epsilon_{0}(t)$ must be time independent, i.e.,
$\epsilon_{0}(t) = \epsilon_{0}$  and similarly $\epsilon_{1}(t) =  \epsilon_{1} - \epsilon_{0} (\alpha^2 -\alpha)/t$.

On the other hand, the small-$s$ expansion can obtained by observing that upon setting on the RHS of (\ref{VII.36a}) $s = \varepsilon^{-1}$  and $sv = v'$ we have
\begin{eqnarray}
\mbox{RHS(\ref{VII.36a})} \ = \ \varepsilon \int_{0}^{\infty}  \rmd v' \ \! e^{-tv'}
\omega(\varepsilon v',t) \delta v(\varepsilon v', t)\, .
\label{VII.38abc}
\end{eqnarray}
If we now use (\ref{VII.36abc}) and expand $\omega(\varepsilon v',t)$ for large
$\varepsilon$  we can easily find that the leading term in the large-$\varepsilon$ behavior of (\ref{VII.38abc}) has the form
\begin{eqnarray}
\mbox{RHS(\ref{VII.36a})} \ \sim \  \epsilon_0 \ \! t^{1-\alpha} \varepsilon^{\alpha -1/2} \frac{\Gamma\left(\alpha -\mbox{$\frac{1}{2}$} \right)}{2\sqrt{\pi}}\, .
\label{VII.39abc}
\end{eqnarray}
By substituting back $\varepsilon = s^{-1}$ we get
%
\begin{eqnarray}
\delta F(s)  \  \sim  \ \epsilon_0 \ \! t^{1-\alpha} \left(\frac{1}{s}\right)^{\!\alpha - 3/2} \frac{\Gamma\left(\alpha -\mbox{$\frac{1}{2}$} \right)}{2\sqrt{\pi}}\, .
\end{eqnarray}
The RHS is $t$ independent only if $\alpha=1$. This also implies that the coefficient $\epsilon_1(t)$ is time independent, i.e., $\epsilon_1(t) = \epsilon_1$  (cf. Eq.~(\ref{V.31.ab})).

With this preparatory analysis we can now substantially simplify  (\ref{VII.37abcd}). In particular, we can write
\begin{eqnarray}
\mbox{\hspace{-5mm}}\delta F(s)
  \ &=& \  \epsilon_0 \frac{\sqrt{s}}{2}\nonumber \\[2mm]
    &+& \  \frac{\epsilon_1}{4} \left(1 + \frac{1}{\sqrt{s}t}  \right) \nonumber \\[2mm]
    &+& \  \frac{\epsilon_2(t)}{8} \left(\frac{1}{\sqrt{s}} + \frac{3}{s t} + \frac{3}{s^{3/2} t^2} \right)\nonumber \\[2mm]
    &+& \  \epsilon_3(t)\left( \frac{1}{s} + \frac{12}{s^{3/2}t} + \frac{60}{s^2 t^2} + \frac{120}{s^{5/2} t^3}    \right)\nonumber \\
   && \vdots \nonumber \\[2mm]
   &=& \  \epsilon_0 \frac{\sqrt{s}}{2} + \frac{\epsilon_1}{4} + \frac{1}{\sqrt{s}}\left(\frac{\epsilon_1}{4t} + \frac{\epsilon_2(t)}{8}  \right) + \ldots \, .
\label{VII.40abc}
\end{eqnarray}
So perturbatively (i.e., order by order in the expansions of $\delta v$ and $\delta F$)
the LHS can equal to the RHS only when $\epsilon_0$ is non-zero and all other $\epsilon_i$ are zero. When we wish to include also
other $\epsilon_i$'s apart from  $\epsilon_0$ then we must include {\em all} of them
in order to allow for mutual compensations of their $t$ dependencies. So for instance, we should chose
$\epsilon_2(t) = \epsilon_2 - 2\epsilon_1/t$ in order to get rid of a time dependence in the term $1/\sqrt{s}$.  The time dependence in the term with $1/s$ will be canceled by $\epsilon_3(t)$, while the time dependence in $1/s^{3/2}$ can be canceled through $\epsilon_3(t)$ and $\epsilon_4(t)$.
In this way one may proceed {\em at infinitum}.

To gain insight into implications of the series expansion (\ref{VII.40abc}), one may assume (e.g., on convergence ground) that the surviving constant coefficients $\epsilon_i$ are decreasing functions of their order ``$i$".
One may then hope that a truncation at a suitable higher order term might give analytically manageable and fairly precise form of $\delta F(s)$. In view of Section~\ref{sec5}, a particularly pertinent truncation is a truncation that terminates
after the $\epsilon_1$ term. In this way we include a first non-trivial contribution beyond $\epsilon_0$. Note also, that such a truncation must include a part from the $\epsilon_2(t)$ term in order to cancel the unwanted $t$ dependence.
%
%
The resulting expansion reads
\begin{eqnarray}
\delta F(s) &=& \  \epsilon_0 \frac{\sqrt{s}}{2} + \frac{\epsilon_1}{4} + \frac{1}{\sqrt{s}}\frac{\epsilon_2}{8}\, .
\label{VII.41abc}
\end{eqnarray}
To a linear order in $\epsilon_i$'s we can write this equivalently as
\begin{eqnarray}
\bar{F}(s) \ &=& \ F(s) + \delta F(s)\nonumber \\
&\approx& \  \frac{\epsilon_1}{4} + \left(1 + \frac{\epsilon_0}{2}\right)\sqrt{s + \frac{\epsilon_2}{4}}\, .
\label{VII.41abcd}
\end{eqnarray}
Here the use was made of the fact that $F(s) = \sqrt{s}$.
In order to ensure that $F(0) = 0$ we should chose $\epsilon_ 1$ so that
\begin{eqnarray}
\epsilon_ 1 \ = \ -2 \left(1 + \frac{\epsilon_0}{2}\right){\sqrt{\epsilon_2}}\, .
\end{eqnarray}

Let us now turn our attention to the analysis of the result (\ref{VII.41abcd}). We begin with the observation that the inverse Laplace transform (see Eqs.~(\ref{II4a}) and (\ref{II6a})) gives us $\hat{\omega}(v,t)$ in the explicit form
\begin{eqnarray}
&&\mbox{\hspace{-10mm}}\hat{\omega}(v,t) \ =\
{\omega}(v,t)  + \delta \omega \nonumber \\[2mm]
&&\mbox{\hspace{-8mm}}= \ \frac{\exp\left\{-\frac{t}{4v}\left[(1+\epsilon_0/2) - v \sqrt{\epsilon_2}  \,\right]^{2}\right\}(1+\epsilon_0/2)}{2\sqrt{\pi} \sqrt{v^3/t}}
\, .
\end{eqnarray}
%
In terms of the mass-smearing distribution this corresponds to the PDF
\begin{eqnarray}
f_{\frac{1}{2}}\!\left(\tilde{m}, tc^2 (1+ \epsilon_0/2)^2, tm^2c^2 + \frac{t\epsilon_2}{4c^2}\right)\, ,
\end{eqnarray}
and to the associated emergent Hamiltonian
\begin{eqnarray}
\bar{H} \ = \ \frac{\epsilon_1}{4} + \left(1 + \frac{\epsilon_0}{2}\right)\sqrt{{{\bi p}^2 c^2 + m^2 c^4  + \frac{\epsilon_2}{4}}}\; .
\label{VII50ab}
\end{eqnarray}
The ensuing  superstatistics PI identity then reads
\begin{widetext}
\begin{eqnarray}
 &&\mbox{\hspace{-10mm}}\int_{\footnotesize{\bi x}(0)\ \! = \ \! \footnotesize{\bi x}'}^{\footnotesize{\bi
x}(t) \ \!  = \ \! \footnotesize{\bi x}} \ \!{\mathcal{D}}{\bi x} \frac{\mathcal{D}
{\bi p}}{(2\pi)^D} \ \! \exp\left\{\int_{0}^{t} \!\!\rmd \tau \
\!\left[\rmi {\bi p}\cdot \dot{\bi x} \ - \
\frac{\epsilon_1}{4} - \left(1 + \frac{\epsilon_0}{2}\right)\sqrt{{{\bi p}^2 c^2 + m^2 c^4  + \frac{\epsilon_2}{4}}}\ \,\right]\right\}
\nonumber \\[2mm]
&&\mbox{\hspace{-1mm}}= \  \int_{0}^{\infty}\!\!\rmd \tilde{m} \
\! f_{\frac{1}{2}}\!\left(\tilde{m}, tc^2 (1+ \epsilon_0/2)^2, tm^2c^2 + \frac{t\epsilon_2}{4c^2}\right)
\int_{\footnotesize{\bi x}(0)\ \! = \ \! \footnotesize{\bi x}'}^{\footnotesize{\bi
x}(t) \ \! = \ \! \footnotesize{\bi x}} \ \!{\mathcal{D}}{\bi x} \ \! \frac{\mathcal{D} {\bi
p}}{(2\pi)^D} \ \! \exp\left\{\int_{0}^{t} \!\!\rmd \tau\
\!\left[\rmi {\bi p}\cdot \dot{\bi x} -  \frac{{\bi p}^2}{2
\tilde{m}}  -  E_0 \right]\right\} .
\label{VII50abcd}
\end{eqnarray}
\end{widetext}
Here
\begin{eqnarray}
{E}_0  \ = \  \left(1 + \frac{\epsilon_0}{2}\right)\left(\sqrt{{m^2 c^4  + \frac{\epsilon_2}{4}}} - \frac{\sqrt{\epsilon_2}}{2}\, \right),
\end{eqnarray}
is the particle's rest energy implied by $\bar{H}$.

In passing we may note that perturbatively  we cannot go beyond a simple re-scaling of the emergent Hamiltonian
because in such a case only $\epsilon_0$ coefficient is non-trivial. In this latter situation the mass-smearing
distribution corresponds to the PDF
\begin{eqnarray}
f_{\frac{1}{2}}(\tilde{m}, tc^2 (1+ \epsilon_0/2)^2, tm^2c^2)\, ,
\end{eqnarray}
and the emergent Hamiltonian has then the form
\begin{eqnarray}
\bar{H} \ = \ \left(1 + {\epsilon_0}/{2}\right)\sqrt{{{\bi p}^2 c^2 + m^2 c^4}}\, .
\label{VII50abd1}
\end{eqnarray}
For the future reference this can be cast in the form
\begin{eqnarray}
\bar{H} \ = \ \sqrt{{{\bi p}^2 \bar{c}^2 + \bar{m}^2 \bar{c}^4}}\, ,
\label{VII51abd2}
\end{eqnarray}
with $\bar{c}^2 = c^2 (1+\epsilon_0/2)^2$ and $\bar{m}^2 = m^2/(1+\epsilon_0/2)^2$.

As a final remark we mention that the key PI identity (\ref{VII50abcd}) alongside with the ensuing
emergent Hamiltonians (\ref{VII50ab}) and (\ref{VII50abd1}) (or (\ref{VII51abd2})) can be nicely
fitted into the framework of doubly special relativity. We will return in more detail to this issue in Section~\ref{sec5}.
\section{The r\^{o}le of gauge fixing in emergent Special Relativity\label{sec6ab}}

As we have mentioned in Section~\ref{sec4} [cf. also Appendices A and B]
the superstatistics PI identity~(\ref{22a}) implicitly corresponds to a special choice of a gauge,
namely to the so called Polyakov or the proper-time gauge. A legitimate question to ask is: what would happen
if a different gauge choice is made? After all, a different gauge fixing condition can change
the physically preferred foliation of spacetime that is central in our ``polycrystalline" picture.
But one could equally well turn this question around and ask whether our granular spacetime with
its preferred foliation could not be the fundamental (or primitive) concept and the
reparametrization invariance only a spurious symmetry related to an inherent redundancy in our description.
%
We shall see in a moment that one may indeed introduce a new redundant variable
into the PI on the RHS of~(\ref{22a}), in such a way that the new action will have the
reparametrization symmetry, but will be still
dynamically equivalent to the original action. By not knowing the source,
one may then view this artificial gauge invariance
as being fundamental or even defining property of the relativistic theory.
One might, however, equally well, proclaim the
``polycrystalline" picture as being a basic (or primitive) edifice of SR and view
the reparametrization symmetry as a mere artefact of an artificial redundancy that
is allowed in our description. It is this second view that we favor
in this paper.
%
%

The trick which will help us to introduce a reparametrization symmetry into the
superstatistics PI~(\ref{22a})
is akin  to the St\"{u}ckelberg
mechanism, whereby one adds a fictitious field to a given system in order to
reveal some hidden properties it might possess~\cite{stuck:38aa,Goto:67a,Slavnov:72a}.
%
For instance, in quantum electrodynamics one can install gauge
symmetry artificially with additional scalar fields, in order to pass from
Proca's ill defined massive Abelian gauge-field theory to renormalizable and gauge
invariant massive electromagnetism. Similar field-enlarging transformations are
also found useful in non-Abelian Yang--Mills theories~\cite{Ruegg:04a} or in
Gravity~\cite{Hinterbichler:11a}.


To proceed, let us introduce a new scalar $\eta$ by making the replacement
%
%
\begin{eqnarray}
{\bi x} \ \mapsto \ {\bi x} + \eta {\bi p}\,  ,
\label{VI49}
\end{eqnarray}
%
which, according to the S\"{u}teckelberg prescription, should follow the pattern of the gauge symmetry we
want to introduce. In addition, in order to keep the boundary conditions for ${\bi x}$
we must require $\eta(0) = \eta(t)$.
Substituting (\ref{VI49}) into the ``non-relativistic Lagrangian" in (\ref{22a}) gives
\begin{eqnarray}
L \ &=& \ \rmi  {\bi p}\cdot\dot{\bi x} \ - \ \frac{1}{2}\left(\frac{1}{\tilde{m}} -
\rmi \dot{\eta}   \right)({\bi p}^2 + m^2c^2)\nonumber \\[2mm]
&& + \ \frac{(\tilde{m} - m)^2 c^2}{2\tilde{m}} \ - \ \frac{\tilde{m} c^2}{2}\, .
\label{VI50a}
\end{eqnarray}
Here we have neglected total derivative terms. Note that the last two
terms can be assimilated into a smearing distribution provided we make the redefinition
\begin{eqnarray}
f_{\frac{1}{2}}\!\left(\tilde{m}, tc^2,
tc^2m^2\right) \mapsto f_{\frac{1}{2}}\!\left(\tilde{m}, tc^2, 0\right)\, .
\end{eqnarray}
Let us further observe that,
\begin{eqnarray}
\delta L  &=&   \frac{\delta L}{\delta x^i(\tau)}\ \! \delta x^i(\tau)  +
 \frac{\delta L}{\delta \eta(\tau)} \ \! \delta \eta (\tau)  +  \frac{\delta L}{\delta p_i(\tau)} \ \!
 \delta p_i(\tau) \nonumber \\[2mm]
&=&   \int \rmd \tau' \! \frac{\delta L}{\delta x'^k(\tau')}
\left(\!\frac{\delta x'^k(\tau')}{\delta x^i(\tau)}\ \! \delta x^i(\tau) +
\frac{\delta x'^k(\tau')}{\delta \eta(\tau)}\ \! \delta \eta (\tau) \! \right)\nonumber \\[2mm]
&&+ \  \frac{\delta L}{\delta p_i(\tau)} \ \! \delta p_i(\tau)\, ,
\end{eqnarray}
and hence the Lagrangian (\ref{VI50a}) is invariant with respect to the gauge transformation
\begin{eqnarray}
\delta {\bi x}(\tau)  \ &=& \  \Lambda(\tau) {\bi p}(\tau)\, ,\nonumber \\
\delta \eta(\tau) \ &=& \  - \Lambda(\tau)\, ,\nonumber \\
\delta {\bi p}(\tau) \ &=& \  {\bi 0}\, ,
\label{VI53a}
\end{eqnarray}
where $\Lambda(\tau)$ is an arbitrary function satisfying $\Lambda(0) = \Lambda(t) = 0$.

At this stage we observe that modulo a multiplicative constant the RHS of the superstatistics
identity (\ref{22a})
can be written as
\begin{widetext}
\begin{eqnarray}
\mbox{(\ref{22a})} \ \! \cong \ \! \int_{\eta(0)=0}^{\eta(t)=0} \mathcal{D}\eta
\int_{0}^{\infty}\!\!\rmd \tilde{m} \
\! f_{\frac{1}{2}}\!\left(\tilde{m}, tc^2, 0 \right) \int_{\footnotesize{\bi x}(0)\ \! = \
\! \footnotesize{\bi x}'}^{\footnotesize{\bi
x}(t) \ \! = \ \! \footnotesize{\bi x}}\!\!\!\mathcal{D}{\bi x} \frac{\mathcal{D}
{\bi p}}{(2\pi)^D} \ \! \exp\left\{\int_{0}^{t} \!\!\rmd \tau\
\!\!\!\left[\rmi {\bi p}\cdot \dot{\bi x}  -   \frac{1}{2}\left(\frac{1}{{
\tilde{m}}}- \rmi \dot{\eta}\right)({\bi p}^2 +   m^2 c^2) \right]\right\}\! .
\label{VI54a}
\end{eqnarray}
\end{widetext}
Here we have used the path-integral analogue of the integral identity
\begin{eqnarray}
\int \! \rmd \eta \rmd x \rmd p \ \! f(x + \eta p, p) \ = \  \mathcal{N} \int \! \rmd x \rmd p \
\! f(x, p)\, ,
\end{eqnarray}
with the constant $\mathcal{N} \equiv \int  \! \rmd \eta$.

Let us now define
\begin{eqnarray}
e(\tau) \ = \ \frac{1}{\tilde{m}} \ - \ \rmi \dot{\eta}(\tau)\, .
\label{VI55a}
\end{eqnarray}
%
This trades the the gauge transformation $\delta \eta(\tau)$ for $\delta e(\tau) = i\dot{\Lambda}(\tau)$.
Note further that $e(t)$ fulfils
the constraint
\begin{eqnarray}
\int_{0}^t \rmd\tau \ \! e(\tau)\ = \ \frac{t}{\tilde{m}}\, ,
\end{eqnarray}
which should be included into a function measure for $e(\tau)$ variables in the form of a delta function.
The equation (\ref{VI54a}) can be then equivalently written as
\begin{widetext}
\begin{eqnarray}
&&\int \mathcal{D}e \int_{0}^{\infty}\!\!\rmd \tilde{m} \ \delta\left(\int_{0}^t \rmd \tau \ \! e(\tau) -
\frac{t}{\tilde{m}} \right)
\! f_{\frac{1}{2}}\!\left(\tilde{m}, tc^2, 0 \right)  \int_{\footnotesize{\bi x}(0)\ \! = \ \!
\footnotesize{\bi x}'}^{\footnotesize{\bi
x}(t) \ \! = \ \! \footnotesize{\bi x}}\!\mathcal{D}{\bi x} \frac{\mathcal{D}
{\bi p}}{(2\pi)^D} \ \! \exp\left\{\int_{0}^{t} \!\!\rmd \tau\
\!\!\!\left[\rmi {\bi p}\cdot \dot{\bi x}  -   \frac{e}{2} ({\bi p}^2 +   m^2 c^2) \right]\right\}
\nonumber \\[3mm]
&&= \ \int \mathcal{D}e \ \! \sqrt{\frac{c^2 t^2 }{2\pi L^3}} \ \!\exp\left(-\frac{t^2c^2}{2L}\right)
\int_{\footnotesize{\bi x}(0)\ \! = \ \! \footnotesize{\bi x}'}^{\footnotesize{\bi
x}(t) \ \! = \ \! \footnotesize{\bi x}}\!\mathcal{D}{\bi x} \frac{\mathcal{D}
{\bi p}}{(2\pi)^D} \ \! \exp\left\{\int_{0}^{t} \!\!\rmd \tau\
\!\!\!\left[\rmi {\bi p}\cdot \dot{\bi x}  -   \frac{e}{2} ({\bi p}^2 +   m^2 c^2) \right]\right\}\! .
\label{VI57a}
\end{eqnarray}
\end{widetext}
Here
In particular,  the total length of particle's trajectory is
\begin{eqnarray}
L \ \equiv \ \int_{0}^{t} \rmd \tau e(\tau)\, .
\end{eqnarray}
In connection with (\ref{VI57a}) we should mention two important points. First, the gauge invariance
of the action can be used to reparametrize the time. Indeed, let $\tau \mapsto \tau + \xi(\tau) \equiv
\lambda(\tau)$ such that $\xi(0) = \xi(t) = 0$. In this case the ``action" takes the form
\begin{eqnarray}
\int_{\lambda_1}^{\lambda_2} \rmd \lambda \left(\rmi {\bi p}\cdot \frac{\rmd {\bi x}}{ \rmd \lambda}  -
\frac{\tilde{e}}{2} \ \! ({\bi p}^2 +   m^2 c^2)\right)\! ,
\end{eqnarray}
where
\begin{eqnarray}
\tilde{e}(\lambda) \ = \ e(\tau) \frac{\rmd \tau}{\rmd \lambda},
\end{eqnarray}
that is, it transforms as the {\em einbein} (i.e., a
square root of the intrinsic metric along the worldline). In particular,
the infinitesimal form of the previous transformation reads
\begin{eqnarray}
\delta e \ = \ - \frac{\rmd (e \xi)}{\rmd \tau}\, .
\end{eqnarray}
This change can be however assimilated into to gauge transformations of ${\bi x}$ and ${\bi p}$
that have the form (cf. (\ref{VI53a}) and (\ref{VI55a}))
\begin{eqnarray}
\delta {\bi x} \ &=& \ \rmi (e \xi) {\bi p}\, ,\nonumber \\[0mm]
\delta {\bi p} \ &=& \  {\bi 0}\, .
\end{eqnarray}
So Eq.(\ref{VI57a}) can be equivalently written as
\begin{widetext}
\begin{eqnarray}
&&\int \mathcal{D}e \ \! \sqrt{\frac{c^2 t^2 }{2\pi L^3}} \ \!\exp\left(-\frac{t^2c^2}{2L}\right)
\int_{\footnotesize{\bi x}(\lambda_1)\ \! = \ \! \footnotesize{\bi x}'}^{\footnotesize{\bi
x}(\lambda_2) \ \! = \ \! \footnotesize{\bi x}}\!\mathcal{D}{\bi x} \frac{\mathcal{D}
{\bi p}}{(2\pi)^D} \ \! \exp\left\{\int_{\lambda_1}^{\lambda_2} \!\!\rmd \lambda \
\!\left[\rmi {\bi p}\cdot \dot{\bi x}  -   \frac{e}{2} ({\bi p}^2 +   m^2 c^2) \right]\right\}
\nonumber \\[3mm]
&& = \ - \frac{~\partial}{c \ \! \partial t} \int \mathcal{D}e \int_{\footnotesize{ x^{\mu}}(\lambda_1)\
 \! = \ \! \footnotesize{x^\mu}'}^{\footnotesize{
x^{\mu}}(\lambda_2) \ \! = \ \! \footnotesize{x^{\mu}}}\!\mathcal{D}{x^{\mu}} \frac{\mathcal{D}
{p_{\mu}}}{(2\pi)^{D+1}} \ \! \exp\left\{\int_{\lambda_1}^{\lambda_2} \!\!\rmd \lambda \
\!\left[\rmi {p}_{\mu}\dot{x}^{\mu}  -   \frac{e}{2} (p^2 +   m^2 c^2) \right]\right\}\! .
\label{IV63a}
\end{eqnarray}
\end{widetext}
In the second line we have utilized an auxiliary Gaussian path integral for $x_0$ in the form~\cite{JK2,PI}
\begin{eqnarray}
&&\frac{\partial}{\partial t}  \int_{\footnotesize{x_0(\lambda_1)}\ \! = \ \!
\footnotesize{0}}^{\footnotesize{x_0(\lambda_2)} \ \! = \ \! \footnotesize{ct}} \!\!\!
\mathcal{D}{x_0} \frac{\mathcal{D}
p_0}{(2\pi)} \ \! \exp\left\{\int_{\lambda_1}^{\lambda_2} \!\!\rmd \lambda \
\!\left[\rmi {p}_{0}\ \! \dot{x}_0  -   \frac{e}{2} \ \! p_0^{2} \right]\right\}\nonumber \\[2mm]
&& = \ -c^2 t \sqrt{\frac{1}{2\pi L^3}} \ \! \exp\left(-\frac{c^2 t}{2L}\right)\! .
\end{eqnarray}
The path integral after the time derivative in (\ref{IV63a}) is already formulated in a covariant way,
as it should be according to Appendix~A. At the same time it is well known that this covariant
path integral is not completely right (see, e.g., Refs.~\cite{PI,polyakov:87}). In fact, it contains an
enormous overcounting, because configurations $(e, x^{\mu}, p_{\mu})$ and
$(e', x'^{\mu}, p'_{\mu})$, that are related to one another by the
gauge transformation (\ref{VI53a}), represent the same
physical configuration. One should use any of the standard techniques of constrained
quantization, such as, e.g., Faddeev--Popov procedure, to remove this redundancy by
imposing appropriately a gauge-fixing condition.
For instance, by utilizing the Polyakov (or proper-time) gauge
\begin{eqnarray}
\dot{e}  \ = \ 0\, ,
\end{eqnarray}
the Faddeev--Popov procedure allows to recast (\ref{IV63a}) into the form
\begin{widetext}
\begin{eqnarray}
\mbox{(\ref{IV63a})} \ = \ - \frac{~\partial}{\partial x_0} \int_0^{\infty} \rmd L \ \!
\int_{\footnotesize{ x^{\mu}}(\lambda_1)\ \! = \ \! \footnotesize{x^\mu}'}^{\footnotesize{
x^{\mu}}(\lambda_2) \ \! = \ \! \footnotesize{x^{\mu}}}\!\mathcal{D}{x^{\mu}} \frac{\mathcal{D}
{p_{\mu}}}{(2\pi)^{D+1}}\ \! \exp\left\{\int_{0}^{L} \!\!\rmd \lambda \
\!\left[\rmi {p}_{\mu}\dot{x}^{\mu}  -   \frac{e}{2} (p^2 +   m^2 c^2) \right]\right\}\! .
\label{IV67ab}
\end{eqnarray}
\end{widetext}
The expression after  $\partial_{x_0}$  is nothing but the well known Feynman--Fock
world-line representation of the KG propagator~\cite{JK2,FH,PI,polyakov:87}.
In addition, in this particular gauge the PI form (\ref{IV67ab}) explicitly coincides
with the superstatistics PI (\ref{22a}) as the reader can easily observe by performing
the $x_0$ and $p_0$ functional integrations.


From the previous considerations we see that the St\"{u}ckelberg trick is
a terrific illustration of the fact that the reparametrization invariance of a
quantum relativistic particle can be regarded as a mathematical sham.
It represents nothing more than a redundancy
of description. In practice one could take any theory and form from it a gauge theory by introducing
redundant variables along the presented lines. Conversely, given any gauge theory, one can always eliminate
the gauge symmetry by eliminating the redundant degrees of freedom. The drawback is that
removing the redundancy is not always a smart thing to do. In fact, it is often said
that gauge symmetry is fundamental as, for instance, in electromagnetism.
A more accurate statement is, however, that the gauge symmetry in electromagnetism is
necessary only if one demands the convenience of {\em linearly} realized LS and
locality. Fixing a gauge will not change the physics, but the price paid is that the LS
and locality are not necessarily manifest. This is precisely what has happened in the case
of the FV equation. There, the particular choice of a gauge in which the equation is formulated leads
to a non-linear realization of the LS (see Appendix~A in Ref.~\cite{JK2} where the associated non-linear
realization is discussed).

In conclusion, the previous construction shows clearly that the polycrystalline spacetime picture
may be legitimately considered as the primitive conceptual framework for a relativistic quantum particle.
On the other hand, the assumptions like the reparametrization invariance, which lie at the bedrock of
relativistic quantum mechanics, cannot be held immune to scrutiny. In fact, we have seen that the reparametrization symmetry, can be perceived as a mere artefact of an underlying (spurious) redundancy of the description.
In other words, the reparametrization symmetry can be seen as being derived, rather than primitive edifice
of relativistic quantum mechanics.

In what follows we propose an unifying approach for Special and Doubly Special Relativity,
based on  the existing laws of quantum mechanics as formulated through
superstatistics PI's.
%
%

\section{Emergent Doubly Special Relativity\label{sec5}}
%
Our analysis from Section~\ref{Sec.Vabc}
reveals what kind of dynamical systems should be expected when
the underlying mass-smearing (or equivalently grain-size) distribution
is slightly deformed. Under a fairly general set of assumptions, we have arrived there at
the emergent Hamiltonian (\ref{VII50ab}) and its ``contracted" version
(\ref{VII50abd1}). In this section we are going to see in more detail, that the dynamics associated with
the aforementioned Hamiltonians  can be identified with the so-called Doubly (or Deformed)
Special-Relativistic dynamics.

In a nutshell, DSR is a theory which coherently
tries to implement a second invariant, besides the speed of light,
into the transformations among inertial reference frames. This new
invariant comes directly from the research in quantum gravity, and
it is usually assumed to be an observer-independent length-scale ---
the Planck length $\ell_p$, or its inverse, i.e., the Planck energy
$E_p = c\ell_p^{-1}$. Thus, it is not
so surprising that the relations mainly studied are those between DSR
and various quantum gravity models~\cite{Smolin, GirOr1, Rovelli}.
In a particularly suggestive approach~\cite{GirOr2},
DSR has been presented as the low energy limit of
Quantum Gravity. Connections between DSR and other theories
(non commutative geometry, AdS space-time, etc.) have also been
recently investigated~\cite{GirLiv}.


It is by now well acknowledged that the DSR stands among the prominent ideas
introduced in physics during the
last decade, and also among the
most controversial ones. Many foundational issues about this theory
are still being debated, in particular for example the
multi-particle sector of the theory (the so called Soccer Ball
Problem)~\cite{Kowalski}. An important area of investigation has been
that of the relations between DSR and other theoretical construction
of modern physics.

Clearly, the most important connection is the one between DSR and
Special Relativity (SR) itself.
However in literature does not exist any conceptually deeper
elaboration of this connection, apart form the obvious statement
that for energy scales much smaller than $E_p$, DSR should reduce to
conventional SR, with leading corrections of first or higher order
in the ratio of energy scales to $E_p$. In this respect, the findings
of the present paper seem to open up new vistas. In fact, when the microstructure
of space-time is considered, then Special Relativity or DSR seem to emerge
from particular choices of such microstructure itself, and from
a non-relativistic Hamilton (i.e., phase-space) mechanics.



%
%

To extend our reasonings to DSR, we start by considering the
modified invariant, or deformed dispersion relation,
\begin{eqnarray}
\frac{\eta^{ab}p_a p_b}{(1 - \ell_p p_0)^2} \  = \  m^2 c^2\, ,
\label{MSI}
\end{eqnarray}
proposed by Magueijo and Smolin~\cite{Mag,Mag2}. Here $m$ plays the role of
the DSR invariant mass. Assuming a metric signature $(+,-,-,-)$, we can solve (\ref{MSI})
in respect to $p_0$, which essentially coincides with the physical Hamiltonian
$\bar{H} = c p_0$. The latter is the generator of the temporal translations with respect to the
coordinate time $t$. Our starting Hamiltonian is therefore
\begin{eqnarray}
\bar{H} \ = \ c\ \!\frac{{- m^2 c^2 \ell  \mp \sqrt{{\bi p}^2(1-m^2 c^2\ell^2) +
m^2 c^2}}}{1-m^2c^2\ell^2}\, ,
\label{HDSR1}
\end{eqnarray}
which we assume as the transformed Hamiltonian $\bar{H}(p,x) = F(H(p,x))$ entering the
proper PI representation of $\bar{P}(x_b,t_b|x_a,t_a)$, see Eq.~(6) and the comments below
[see also Ref.~\cite{JK1}]. Note that by setting
\begin{eqnarray}
\epsilon_1 \ &=& \ 2 \left(\sqrt{\frac{1}{1-c^2 m^2 \ell^2}}\ \!  - \ \! 1 \right), \nonumber \\[2mm]
\epsilon_2 \ &=& \   \frac{4c^6 m^4 \ell^2}{1-c^2 m^2 \ell^2}\, ,
\end{eqnarray}
we can identify the DSR Hamiltonian
(\ref{HDSR1}) with the Hamiltonian (\ref{VII50ab}) obtained in Section~\ref{Sec.Vabc}.
In close analogy with (\ref{22a}), it is now possible to show the superstatistics identity
[see Appendix C]
\begin{widetext}
\begin{eqnarray}
 &&\mbox{\hspace{-15mm}}\int_{\footnotesize{\bi x}(0)\ \! = \ \!
 \footnotesize{\bi x}'}^{\footnotesize{\bi
x}(t) \ \!  = \ \! \footnotesize{\bi x}} \ \!{\mathcal{D}}{\bi x} \frac{\mathcal{D}
{\bi p}}{(2\pi)^D} \ \! \exp\left\{\int_{0}^{t} \!\!\rmd \tau \
\!\left[\rmi {\bi p}\cdot \dot{\bi x} \ + \
c\frac{\left({m^2 c^2 \ell  - \sqrt{{\bi p}^2(1-m^2c^2\ell^2) + m^2 c^2}}\right )}
{(1-m^2c^2\ell^2)}\right]\right\}
\nonumber \\[2mm]
&&\mbox{\hspace{12mm}}= \  \int_{0}^{\infty}\!\!\rmd \tilde{m} \
\! f_{\frac{1}{2}}\!\left(\tilde{m}, tc^2 \lambda, tc^2m^2 \lambda\right)
\int_{\footnotesize{\bi x}(0)\ \! = \ \! \footnotesize{\bi x}'}^{\footnotesize{\bi
x}(t) \ \! = \ \! \footnotesize{\bi x}} \ \!{\mathcal{D}}{\bi x} \ \! \frac{\mathcal{D} {\bi
p}}{(2\pi)^D} \ \! \exp\left\{\int_{0}^{t} \!\!\rmd \tau\
\!\left[\rmi {\bi p}\cdot \dot{\bi x} -  \frac{{\bi p}^2}{2
\tilde{m}}  -  E_0 \right]\right\} ,
\label{31a}
\end{eqnarray}
\end{widetext}
where $E_0 = m c^2/(1+ m c \ell)$ is the particle's rest energy
[see, e.g., \cite{Mag2}] and $\lambda = 1/(1-m^2c^2\ell^2)$ is the
deformation parameter. From (\ref{23a}), it is easy to see that
$\langle \tilde{m} \rangle = m + 1/(t c^2 \lambda)$ and
var$(\tilde{m}) = m/tc^2\lambda +  2/t^2c^4\lambda^2$.

From the structure of $\langle \tilde{m} \rangle$ we can obtain
further useful insights. Similarly as in the SR framework,
the fluctuating Newtonian mass $\tilde{m}$ converges
rapidly, at long times $t$, to the SR rest mass $m$.
But in this case is the rate of convergence
controlled also by the parameter $\lambda$. Reminding
that $E_p=c/\ell_p$, we see that $\lambda = 1/(1-E^2/E_p^2)$.
So, $\langle \tilde{m} \rangle$ can converge
rapidly to the Einstein value $m$, even at short times,
provided that the particle's energy $E$ be close
to the Planck energy $E_p$. The correlation distance is
now given by $\sim 1/(m c \lambda)$, and since $\lambda > 1$,
then $1/(m c \lambda) < 1/(m c)$ always.

From the identity (\ref{31a}) we can quickly deduce the CCR via the
standard PI analysis. In particular the CCR can be directly related
to the degree of roughness (described through Hausdorff dimension
$d_H$ or Hurst exponent $h$) of typical PI
paths~\cite{Kroger,FH}. For instance, the usual non-relativistic canonical relation
$[\hat{{\bi x}}_i,\hat{{\bi p}}_j] = \rmi\delta_{ij} $ results from the fact that, for a typical
path occurring in non-relativistic PI's, $d_H$ and $h$  are $2$
and $1/2$, respectively. In fact, in non-relativistic quantum mechanics all {\em local} potentials
fall into the same universality class (as for the scaling behavior)
as the free system~\cite{Kroger}. The latter might be viewed as a PI justification of the
universal form of non-relativistic CCR's.

It is not hard to show [cf. Appendix D] that the PI identity (\ref{31a}) implies the commutators
\begin{eqnarray}
[\hat{\bi x}_i,\hat{\bi p}_j]_{\rm DSR1} \ =
\ \rmi\!\left(\delta_{ij} + \frac{\kappa^2 - m^2c^2}{\kappa^2m^2c^2}\
\!\hat{\bi p}_i \hat{\bi p}_j\right)\! .
\label{27aab}
\end{eqnarray}
Here $\kappa = 1/\ell$. The CCR (\ref{27aab}) resembles the Snyder version
of the deformed CCR associated to the dispersion relation (\ref{MSI})
[cf. Refs.~\cite{Snyder:47,Ghosh:07,Ghosh:06,Banerjee:06}].
To be precise, the Snyder fundamental commutation relation [see Ref.~\cite{Snyder:47}]
in the notation of the present paper (for $\hbar = 1$, the Snyder fundamental
length $a \equiv \ell = 1/\kappa$) would read
\begin{eqnarray}
[\hat{\bi x}_i,\hat{\bi p}_j]_{\rm Snyder} \ =
\ \rmi\!\left(\delta_{ij} + \frac{\hat{\bi p}_i \hat{\bi p}_j}{\kappa^2}\ \!\right)\! .
\label{Snyder}
\end{eqnarray}
So the prefactor of the deforming term in the Snyder commutator is a \emph{constant}
related to the fundamental length, while in the DSR commutator (\ref{27aab})
the prefactor varies with the Einstein mass of the particle considered.
The minimal length
interval $\ell$ is typically set to be the Planck length $\ell_p$, or more generally
to be the Compton length $\lambda_C$ (which reduces to the Planck length for a Planck mass).
For definiteness we will in the following identify $\ell$  with $\ell_p$.
In this connection, note that when $mc \to \kappa$, i.e., when $m$ coincides with the Planck mass,
then the CCR (\ref{27aab})
becomes non-relativistic. This can also be directly seen from (\ref{31a}),
where for $m \to M_p$ the defromation parameter $\lambda \to \infty$, and the smearing distribution
$f_{\frac{1}{2}}\!\left(\tilde{m}, tc^2 \lambda, tc^2m^2 \lambda\right) \to
\delta(m-\tilde{m})$, which yields the usual PI for a Wiener process.


Should we have used instead of (\ref{MSI}) a different DSR dispersion relation, for example
\begin{eqnarray}
\frac{p_0^2 - {\bf p}^2}{1 - (\ell_p p_0)^2} \ = \  m^2 c^2\, ,
\end{eqnarray}
(which is discussed in Ref.~\cite{Mag3}), we would have obtained the DSR Hamiltonian
\begin{eqnarray}
\bar{H} \ = \ \pm \ \! \frac{\sqrt{{\bi p}^2c^2 + m^2 c^4}}{\sqrt{1 + m^2 c^2 \ell^2}}\, ,
\label{28bba}
\end{eqnarray}
from which follows the superstatistics identity [see again Appendix C]
\begin{widetext}
\begin{eqnarray}
 &&\mbox{\hspace{-15mm}}\int_{\footnotesize{\bi x}(0)\ \! = \ \! \footnotesize{\bi x}'}^{\footnotesize{\bi
x}(t) \ \!  = \ \! \footnotesize{\bi x}} \ \!{\mathcal{D}}{\bi x} \frac{\mathcal{D}
{\bi p}}{(2\pi)^D} \ \! \exp\left\{\int_{0}^{t} \!\!\rmd \tau \
\!\left[\rmi {\bi p}\cdot \dot{\bi x} \ - \ \frac{\sqrt{{\bi p}^2c^2 + m^2 c^4}}{\sqrt{1 + m^2 c^2 \ell^2}}
\right]\right\} \nonumber \\[2mm]
&&\mbox{\hspace{12mm}}= \  \int_{0}^{\infty}\!\!\rmd \tilde{m} \
\! f_{\frac{1}{2}}\!\left(\tilde{m}, tc^2 {\zeta}^2, tc^2m^2 \right)
\int_{\footnotesize{\bi x}(0)\ \! = \ \! \footnotesize{\bi x}'}^{\footnotesize{\bi
x}(t) \ \! = \ \! \footnotesize{\bi x}} \ \!{\mathcal{D}}{\bi x} \ \! \frac{\mathcal{D} {\bi
p}}{(2\pi)^D} \ \! \exp\left\{\int_{0}^{t} \!\!\rmd \tau\
\!\left[\rmi {\bi p}\cdot \dot{\bi x} -  \frac{{\bi p}^2}{2
\tilde{m}}  -  \bar{E}_0 \right]\right\} .
\label{31aaa}
\end{eqnarray}
\end{widetext}
Here $\bar{E}_0  = mc^2/\sqrt{1 + m^2c^2\ell^2}$ is the particle's rest energy and the
deformation parameter now reads $\zeta = 1/\sqrt{1 + m^2c^2\ell^2}$.
It is not difficult to compute that
$\langle \tilde{m}\rangle = m/{\zeta} + 1/(c^2 t {\zeta}^2)$
and var$(\tilde{m}) = {m}/({tc^2\zeta^3}) + {2}/({t^2c^4\zeta^4})$.
Let us observe that this double-special relativity model does not have the
desired property that its fluctuating mass converges to a Lorentz mass in the
large $t$ limit. In addition, because $\zeta \in (0,1)$, the fluctuations at
short times cannot be suppressed and one thus cannot hope to have a relativistic
system with a sharp Einsteinian mass at the Planck energy.
In passing, we may note that the DSR system  (\ref{28bba}) coincides with system (\ref{VII50abd1})-(\ref{VII51abd2})  from Section~\ref{Sec.Vabc}
provided we make identification
\begin{eqnarray}
\epsilon_0 \ = \ 2\left(\zeta \ \! - \ \! 1 \right).
\end{eqnarray}

The CCR in this system read [cf. Appendix D]
%
\begin{eqnarray}
[\hat{\bi x}_j, \hat{\bi p}_i]_{\rm DSR2} \ &=&
\ \rmi\!\left(\delta_{ij} + 
\!\frac{\hat{\bi p}_i \hat{\bi p}_j}{m^2c^2}\right)\! .
\label{30aab}
\end{eqnarray}
As the reader can see, this commutator coincides with the SR one. Some comments on this
apparently surprising fact are contained in Appendix D.

At this point, for the sake of completeness, we should also note that commutator~(\ref{27aab})
does not coincide with the commutator (61) of the Magueijo--Smolin paper~\cite{Mag2}, although
it comes from the same dispersion relation (\ref{MSI}), proposed in~\cite{Mag2} as formula~(3).
Moreover, both commutators (\ref{27aab}) and (\ref{30aab}) do not enjoy an important property
of the commutator (61) from Ref.~\cite{Mag2}. In fact, when the energy of the boosted system
approaches the Planck energy, the aforementioned commutator, which reads
\begin{eqnarray}
[\hat{\bi x}_j, \hat{\bi p}_i]_{\rm MS} \ &=&
\ \rmi \, \delta_{ij} \left(1 - \frac{E}{E_p}\right)\! ,
\end{eqnarray}
goes to zero, and so it predicts the appearance of a classical world at the Planck scale.
On the contrary, when the energy of the particle under boost approaches the Planck value,
commutators (\ref{27aab}) and (\ref{30aab}) become respectively
%
\begin{eqnarray}
&&[\hat{\bi x}_j, \hat{\bi p}_i]_{\rm{DSR1}} \ = \ \rmi \, \delta_{ij}\, ,\nonumber \\
&&[\hat{\bi x}_j, \hat{\bi p}_i]_{\rm{DSR2}}  \ = \ \rmi \left(\delta_{ij} +
\frac{\hat{\bi p}_i \hat{\bi p}_j}{\kappa^2}\right)\, .
\end{eqnarray}
%
%

The qualitative difference in behavior of both CCR can be traced back to the
fact that commutators in (doubly-)special relativity depend on two things;
First, the fundamental commutators are essentially the Dirac brackets of the canonical variables.
The explicit definition of the Dirac brackets depends on the
choice of a gauge (gauge fixing condition), which for relativistic systems corresponds to choice
of a specific physical time. So the commutation relations are generally gauge fixing dependent
in both SR and DSR systems. Second, the fundamental commutator $[\hat{\bi x}_j, \hat{\bi p}_i]$
depends (through the Jacobi identities) on the whole symplectic structure of the system (and therefore
also on the commutator $[\hat{\bi x}_j, \hat{\bi x}_i]$, for example). These are not specified
by a particular DSR model, but they have to be chosen aside. Of course, one obtains different
theories for different choices of $[\hat{\bi x}_j, \hat{\bi x}_i]$.

In our specific path-integral approach, the gauge which is automatically incorporated
in the path integral is the Polyakov gauge. We obtain the same fundamental commutator
of Ghosh \cite{Ghosh:07}. For Ghosh~\cite{Ghosh:07} and Mignemi~\cite{Mignemi} the
$\hat{\bi x}_j$'s do not commute, $[\hat{\bi x}_j, \hat{\bi x}_i] \neq 0$, while on
the contrary Magueijo--Smolin in~\cite{Mag2} require the $\hat{\bi x}_j$'s to commute (formula (60)).

So, a DSR theory is not only defined by the dispersion relation, but also by the gauge
fixing and the choice of the symplectic structure, which is essentially arbitrary.
In principle, therefore, only the experiment can effectively discriminate among different models.
It should now become
clear why our model can produce commutators different from those
of Ref.~\cite{Mag2}, although both models share the same deformed dispersion relation.
As a small further note, we may add that from the Magueijo--Smolin paper~\cite{Mag2} it is
not clear if the proposed commutators satisfy the Jacobi identities or not (not enough
commutators are, in fact, explicitly specified to enable the reader to verify the Jacobi identity).

{\revision
We conclude with an important observation. Should we have applied our analysis from Section~\ref{Sec.Vabc}
to the above DSR systems we would have obtained that a slight perturbation in the mass-smearing distribution
would yield again DSR systems. From this standpoint is the DSR (as well as its low-energy limit --- SR) a robust concept, i.e. its algebraic structure continues to hold despite (potentially dynamical) alterations in
polycrystalline structure conditions.
}

%
%


\section{Concluding remarks}

In this paper we have shown that both SR and DSR systems can arise by statistically
coarse-graining underlying non-relativistic (Wiener) process, making the latter more
fundamental and the former in some sense emergent. The coarse-graining can be viewed
as arising from superposition of two stochastic processes.
On a short spatial scale (much shorter than particle's Compton wavelength) the particle
moves according to a Brownian, non-relativistic, motion. Its Newtonian mass
fluctuates according to an inverse Gaussian distribution.

The time-compounded smearing distribution tends, however, rapidly to the delta-function
distribution due to the central limit theorem. This happens at the time scale of the order
of Compton time, at which the relative mass fluctuation is of order unity.
The Compton length also represents the critical length scale at which the Feynman--Hibbs
scaling relation between $\Delta x$ and $\Delta t$ changes its critical
exponent (Hurst exponent) from $1$ to $1/2$. The Compton time
and ensuing Compton length can be thus viewed as correlation time and correlation length,
respectively.
The averaged (or coarse-grained) velocity over the correlation time is the light velocity $c$.
On a time scale much larger than the Compton time, the particle then behaves as a
relativistic particle with a sharp mass equal to Einstein's (i.e., Lorentz invariant)
mass. In this case the particle moves with a net velocity which is less than $c$.
Here the reader may notice a close analogy with the Feynman chessboard PI.
In contrast to the chessboard PI, our approach is not confined to only $1+1$ dimensional Dirac fermions.

%



The presented concept of {\em statistical emergence}, which
is shared both by SR and DSR, can offer a new valid insight into the Planck-scale
structure of space-time. The existence of a discrete polycrystalline substrate (or vacuum) might
be welcomed in various quantum gravity constructions.
In fact, it has been speculated for long time that quantum gravity
may lead to a discrete structure of space and time which
can cure classical singularities. This idea has been embodied, in
particular, in Loop Quantum Cosmology~\cite{Ashtekar,Bojowald}.
A similar proposal was put forward in Ref.~\cite{Vilenkin} in connection
with the space-time foam. It should be stressed that many condensed matter systems
show that a discrete sub-structure might lead to a genuinely relativistic dynamics at
low energies~\cite{Volovik03}, without any internal inconsistency.
A paradigmatic example of this are wide single-layer
carbon crystals (graphene), where an effective theory emerges
in which conducting electrons behave, at low temperatures, as massless relativistic
Dirac fermions with a ``light speed" equal to the Fermi velocity of the
crystal~\cite{Novoselov:05}.
Essentially the same emergent behavior is known to hold also for silicene, i.e.,
the monolayer silicon equivalent of graphene~\cite{Vogt:12a}.

%

In this connection one can also stress that crystal-like substrates  --- discrete lattices,
are routinely used, for instance, in computational quantum
field theory~\cite{KleinertIII,Creutz:83,FIRST} where the genuine relativistic field dynamics emerges only in the
long-wavelength limit, i.e., at distances much larger than a typical lattice spacing.
However, with a few notable exceptions~\cite{Friedberg:83a,Friedberg:83b}, the lattices is these case mainly serve as numerical regulators of
ultraviolet divergences. Indeed, a crucial point of renormalized
theories is precisely to extract lattice-independent data
from numerical computations.

We close this paper with a number of questions, to be examined in a future work.
Foremost: We have stressed that the proposed coarse-graining view directly applies
also to simple interacting systems, such as the Dirac's particle in a constant external
electromagnetic potential. One may naturally wonder whether this interpretation can
be extended to more general interacting systems. So far, two points hinder this
program to be carried further in a full generality: first, the general ${\bi x}$-dependence
of $A_{\mu}$ and $V$ leads to a notorious ordering problem. Second, and most important,
the transformation that would bring the Hamiltonian into a form where the positive and
negative energy parts are explicitly separated is no longer possible for a general
interaction. This last point makes it difficult to carry over straightforwardly our
reasonings. On the other hand, if the proposed picture aspires to be more than just
an interesting metaphor,
%
%
one should be able to demonstrate the viability
of our scheme also for less trivial interactions.
It remains yet to be seen to what extent this can be done.

Another open issue is the r\^{o}le of the smearing distribution. In the presented
approach, the specific form of the smearing distribution is a mere byproduct of
the superstatistics PI paradigm. Our heuristic picture of a polycrystalline medium
which we affiliate with the smearing distribution is clearly not the only possibility.
Furthermore, a deeper understanding of a dynamical origin of our smearing
would be highly desirable.
In fact, the exact LS is due to a very special form of the Newtonian-mass distribution.
The exact LS of a spacetime has no fundamental
significance in our model, but it is only an accidental symmetry of the
spatially coarse-grained theory. This ``accidentalness" is controlled by
a specific form of the grain distribution. We have seen that a small
departure from its shape brings a departure from LS and leads naturally to the DSR.
In this respect a useful guide
to understanding the specificity of grain distributions could be the observation
that generalized inverse Gaussian distributions with $p=1/2$
correspond (among others) to first-hitting inverse-time PDF's~\cite{Barndorff:78}.
This might indicate that our smearings distribute inverse times
(i.e., masses) between successive events in a renewal process
(such as a passage to a new grain). The latter would, in a sense,
support our polycrystalline picture.

Finally, it is also hoped that the essence of our results will continue to hold in curved spacetimes.
This could be an important step in addressing the issue of quantum gravity. In this connection
we may notice a conceptual similarity with the Ho\v{r}ava--Lifshitz gravity theory~\cite{Horava}, where,
as in our case, space and time are not equivalent at the fundamental level, and therefore
the theory is intrinsically non-relativistic. The relativistic concept of time together
with its Lorentz invariance emerges at distances much larger than Compton wavelength.

\vspace{2mm}
\section*{Acknowledgments}

A particular thank goes to S.~Mignemi for his enlightening emails on the DSR commutators
and DSR symplectic structure.
We are grateful also for comments from
H.~Kleinert, C.~Schubert, F.~Bastianelli,
and L.S.~Schulman who have helped us to understand better the ideas
proposed in this paper. The figure is due to the art of M.~Nespoli.
P.J. is supported by
the Czech Science Foundation under the Grant No. P402/12/J077.
F.S. is supported by Taiwan National Science Council under Project No. NSC 97-2112-M-002-026-MY3.

 \section*{Appendix A~\label{ap1}}
%
Here we briefly review, without going to much details, some essentials related to PI representation of
the Klein--Gordon particle in the Feshbach--Villars representation.

The doubling of the wave function in Eq.~(\ref{III10a}) implies the simultaneous description of
particles and antiparticles~\cite{Feshbach58}. The Hamiltonian ${H}_{_{\rm FV}}({\bi p})$
can be diagonalized as
\begin{eqnarray}
\hat{H}_{_{\rm FV}} ({\bi p}) \ = \ {U}({\bi p})\sigma_3{U}({\bi p})^{-1}
\hat{H}({\bi p})\, ,
\label{12a}
\end{eqnarray}
where $U$ is non-unitary hermitian matrix
\begin{eqnarray}
\mbox{\hspace{-8mm}}&&{U}({\bi p})  =
\frac{(1+ \gamma_{v}) + (1- \gamma_{v})\sigma_1}{2\sqrt{\gamma_{v}}}\, ,
\end{eqnarray}
$\hat{H}({\bi p}) = c\sqrt{\hat{\bi p}^2 + m^2c^2}$ is the
energy operator, and $\gamma_v = 1/\sqrt{1-v^2/c^2}$. The Green's function ${\mathcal{G}}(x,y)$ associated
with the FV Schr\"{o}dinger equation can be written as
%
%
\begin{widetext}
\begin{eqnarray}
{\mathcal{G}}(x;y) \ = \ \frac{\rmi}{c^2}\int_{\mathbb{R}^4} \frac{\rmd^{D+1} p
}{(2\pi)^{D+1}} \ \! \frac{e^{-\footnotesize{\rmi} p(x-y)}}{p^2 -m^2 c^2 + \rmi
\epsilon} \ \!\left[p_{_0} c + (\sigma_3 + \rmi\sigma_2) \frac{{\bi
p}^2}{2m} +\sigma_3 m c^2 \right]\, .
\label{16a}\nonumber\\[-2mm]
\end{eqnarray}
%
Here the $\rmi \epsilon$ prescription corresponds to the usual
Feynman boundary condition. Note that the imaginary-time Green function
$\mathcal{G}({\bi x},-it;{\bi x}',-it') \equiv P({\bi x},t|{\bi
x}',t')$ is a solution of the Fokker--Planck like equation
$(\partial_t + \hat{H}_{_{\rm FV}}) P({\bi x},t|{\bi x},t') \ = \
\delta(t-t') \delta^{(3)}({\bi x} - {\bi x}')$,
%
where $P({\bi x},t|{\bi x}',t') = \langle {\bi x}| e^{-(t-t')
\hat{H}_{_{\rm FV}}}| {\bi x}'\rangle$. Because of (\ref{12a}), the latter can be equivalently written as
%
%
\begin{eqnarray}
P({\bi x},t|{\bi x}',t') =
 \int_{\mathbb{R}^D} \rmd {\bi x}''  \int_{\mathbb{R}^D}
 \frac{\rmd {\bi p} }{(2\pi)^{D}}\ \!
\ \! {e^{\rmi \small{{\bi p}\ \!\cdot({\bi x} - {\bi x}'')}}}
 \ \! U({\bi p}) \langle{\bi x}''|e^{-(t-t')
\sigma_3\hat{H}({\footnotesize{\bi p}})}|{\bi x}' \rangle U({\bi p})^{-1}\, .
\label{14a}\nonumber\\[-3mm]
\end{eqnarray}
\end{widetext}
At this point a warning should be made, that we cannot write naively the PI representation of
$\langle{\bi x}|e^{-(t-t')\sigma_3 \hat{\! H}({\footnotesize{\bi p}})}|
{\bi x}'\rangle$ by considering  $\sigma_3 \hat{\! H}({\footnotesize{\bi p}})$ as a formal Hamiltonian.
This is because the PI with the ensuing action $\int_{t'}^{t} \rmd \tau\left[\rmi \small{\bi
p}\cdot\dot{\small{\bi x}} - c \sigma_3 \sqrt{\small{\bi p}^2 + m^2 c^2}\right]$
would diverge. The pathology involved can be evaded by forming superpositions of  integrals
which differ for upper and lower components of $\exp(-t\sigma_3
\hat{H}{(\bi p}))$ according to Feynman--Stuckelberg prescription, i.e., we consider positive frequencies
as propagating forward in time and negative frequencies backward in time.
This yields the well behaved PI [cf. Ref.~\cite{JK2}]
\begin{widetext}
\begin{eqnarray}
\langle{\bi x}|e^{-t\sigma_3 \hat{H}({\footnotesize{\bi p}})}| {\bi x}'\rangle \ =
\ \int_{0}^{\infty} \rmd v \ \!\omega(v,t_{})
\int_{\footnotesize{\bi x}(0)\ \! = \ \! \footnotesize{\bi x}'}^{\footnotesize{\bi
x}(t) \ \! = \ \! \footnotesize{\bi x}}\ \!\mathcal{D}{{\bi x}}
\frac{\mathcal{D}{{\bi p}}}{(2\pi)^D} \ \!
 \exp\left\{\int_{0}^{t}\rmd \tau\left[\rmi {\bi p}\cdot\dot{\bi x} -
v ({{\bi p}^2c^2 + m^2 c^4)}\right]\right\}\, .
\label{15a}\nonumber\\[-3mm]
\end{eqnarray}
The weight function is now a matrix valued Weibull distribution
\begin{eqnarray}
\omega(v,t)  \ = \  \frac{1}{2\sqrt{\pi} \sqrt{{v^3}/{|t|}}}
\left(
\begin{array}{cc}
\theta(t) \ \! e^{-t/4v} & 0 \\
0 & \theta(-t)\ \! e^{t/4v} \\
\end{array}
\right)\! .
\label{16c}
\end{eqnarray}
\end{widetext}
%
%
At the same time we can write
\begin{eqnarray}
&&\mbox{\hspace{0mm}}\langle{\bi x}|\rme^{-t\sigma_3 \hat{\! H}({\footnotesize{\bi p}})}| {\bi x}'\rangle
\ = \ {\theta(t)} \!\frac{1 + \sigma_3}{2} \ \!
 \langle{\bi x}|\rme^{-t
\hat{\! H}({\footnotesize{\bi p}})}| {\bi x}'\rangle \nonumber \\
&&\mbox{\hspace{26mm}}+ \ {\theta(-t)}\ \!\frac{1 -
\sigma_3}{2} \ \!\langle{\bi x}|\rme^{t \hat{\! H}({\footnotesize{\bi p}})}| {\bi
x}'\rangle \nonumber \\[2mm]
&&\mbox{\hspace{10mm}}= \ \frac{1}{2}\left(1  - \frac{ \hat{\! H}({\footnotesize{\bi p}})\ \!\sigma_3 }{
\partial_t}\right) \langle{\bi x}|\rme^{-|t| \hat{\! H}({\footnotesize{\bi p}})}| {\bi
x}'\rangle   \, . \label{85ab}
\end{eqnarray}

Comparing the Euclidean version of
${\mathcal{G}}(x,y)$ represented by Eq.~(\ref{16a}) with  (\ref{14a}) and (\ref{85ab})
we see that $\langle{\bi x}|\rme^{-t \hat{\! H}({\footnotesize{\bi p}})}| {\bi
x}'\rangle $ (with positive $t$) can be written as a time derivative of a covariant quantity,
namely the KG propagator. It can be shown (see~\cite{JK2}) that the resulting covariant quantity is a PI
which coincides with the familiar Feynman--Fock's representation of the KG propagator in the
Polyakov (or proper-time) gauge~\cite{PI,polyakov:87}.

Because of the formal similarity of
the diagonalization (\ref{12a}) with a Foldy--Wouthuysen diagonalization~\cite{Foldy50},
one can treat spin-$\frac{1}{2}$ fermions in close analogy with
KG particles [see~\cite{JK2} for details]. {\revision It should be also noted that Foldy--Wouthuysen-like
diagonalizations are quite standard
also for particles with a higher spin. This is particularly clearly seen when the higher-spin particle wave equations are phrased via Bargmann--Wigner equations~\cite{Ohnuki}. There the corresponding wave functions, the so-called Bargmann--Wigner amplitudes, can
be again transformed into form where the positive and negative frequency parts are explicitly separated.
For these reasons our superstatistics formula (\ref{15a}) will still hold with the alteration that for a spin $n/2$ particle the smearing matrix will be $2^{n+1}\!\times 2^{n+1}$ matrix.  }
%
%

\section*{Appendix B~\label{apb}}

We show here how to obtain the CCR for the non relativistic
and special relativistic systems discussed in the main text. Doubly-special-relativistic
systems are then considered in Appendix D.\\

{\em a) Non relativistic systems.} --- From the invariance of non-relativistic PI's with respect
to translation by a fixed path we obtain the Ward identity in the form~\cite{schulman}
\begin{eqnarray}
\!\!\!\!\langle {\bi x}'',t''| \frac{i \delta_{ij}}{\tilde{m}} +
[\dot{\hat{\bi x}}_i, \hat{\bi x}_j]|_{\tau}|{\bi x}',t'\rangle \ =
\ 0\, , \,\,\,\,\,\,\, t'< \tau <  t''\, .
\label{31aa}
\end{eqnarray}
The completeness of the Heisenberg base vectors
$\{| {\bi x},t \rangle \}$ turns the {\em weak} matrix relation (\ref{31aa})
into a {\em strong} operatorial identity ---  the usual non relativistic CCR.
For analytically continued PI (such as the PI (\ref{2a})) this reads
\begin{eqnarray}
\langle {\bi x}'',t''| \frac{\delta_{ij}}{\tilde{m}} +
[\dot{\hat{\bi x}}_i, \hat{\bi x}_j]|_{\tau}|{\bi x}',t'\rangle \ =
\ 0\, , \,\,\,\,\,\,\,\,\,t'< \tau <  t''\, .~~~
\label{32aa}
\end{eqnarray}

The conclusions (\ref{31aa}) and (\ref{32aa}) are true for any non-relativistic system
with a kinetic term of the form $m \dot{\bi x}^2/2$. It can be shown~\cite{schulman,Kroger}
that the previous Ward identities imply the Feynman--Hibbs scaling~\cite{FH}
\begin{eqnarray}
\langle {\bi x}'',t''|\! \ |\hat{\bi x}_i(\tau + \Delta t) - \hat{\bi x}_i(\tau)|\! \ |{\bi x}',t'\rangle
\ \propto \ (\Delta t)^{1/2}\, .
\label{B.34}
\end{eqnarray}
This means that the Hurst exponent $h$ of a representative trajectory is $1/2$ and
the corresponding Hausdorff fractal dimension $d_H$ is $2$. In this respect the
non-relativistic PI trajectories are reminiscent of a Wiener process.\\

{\em b) Special Relativity.} --- In the relativistic framework the Ward identity (\ref{32aa})
boils down to [cf. identities (\ref{22a}) and (\ref{25a})]
\begin{eqnarray}
\mbox{\hspace{-5mm}}&&\int_{0}^{\infty}\!\!\rmd \tilde{m} \
\! f_{\frac{1}{2}}\!\left(\tilde{m}, tc^2, tc^2m^2 \right)
\langle {\bi x}'',t''| \frac{\delta_{ij}}{\tilde{m}} +
[\dot{\hat{\bi x}}_i, \hat{\bi x}_j]|_\tau |{\bi x}',t'\rangle \nonumber \\
\mbox{\hspace{-5mm}}&&\,\,\, = \ 0\, ,
\label{35abb}
\end{eqnarray}
where $t = t'' - t'$.

In order to explicitly compute the previous integral we need to recall several formulae.
First, since the {\em base-vectors} in Heisenberg picture are time dependent
while in Schr\"{o}dinger picture they are time independent (in contrast to
{\em state-vectors}), we can write
%
\begin{eqnarray}
\mbox{\hspace{-4mm}}|{\bi x}',t'\rangle \ &=& \ e^{t' \hat{H}_C}\,\,|{\bi x}'\rangle\, , \nonumber\\
\mbox{\hspace{-4mm}} \langle{\bi x}'',t''| \ &=& \ \langle{\bi x}''|\,\,e^{-t'' \hat{H}_C}\, ,\nonumber\\
\mbox{\hspace{-4mm}} [\dot{\hat{\bi x}}_i, \hat{\bi x}_j]|_{\tau} \
 &=& \ e^{\tau \hat{H}_C}\,\, [\dot{\hat{\bi x}}_i, \hat{\bi x}_j] \,\,e^{-\tau \hat{H}_C}\, ,
\label{evol}
\end{eqnarray}
where $H_C$ is the classical, i.e., non relativistic Hamiltonian. A second relation useful in the
simplification of (\ref{35abb}) is formula (6) of Ref.~\cite{JK1}, namely
%
%
for Hamiltonians not explicitly dependent on time we have
\begin{eqnarray}
\mbox{\hspace{-2mm}}e^{-t\, H_{SR}} \ = \ \int_{0}^{\infty}\rmd \tilde{m} \
\! f_{\frac{1}{2}}\!\left(\tilde{m}, t\, c^2,  t\, c^2m^2\right)
e^{- t\, H_C}\,  ,
\label{49}
\end{eqnarray}
with
\begin{eqnarray}\nonumber \\[-9mm]
H_{SR} &=& c\sqrt{{\bi p}^2 + m^2 c^2}\, , \nonumber \\
H_C &=& \frac{{\bi p}^2}{2\tilde{m}} \ + \  m c^2 \, .\\[-1mm]
\nonumber
\end{eqnarray}
Combining (\ref{35abb}) and (\ref{49}) with yet another identity, namely
\begin{eqnarray}
\mbox{\hspace{-2mm}}\frac{e^{-t\, H_{SR}}}{m\,\gamma_v} \ =  \ \int_{0}^{\infty}\rmd \tilde{m}
\ f_{\frac{1}{2}}\!\left(\tilde{m},  t\, c^2,  t\, c^2m^2\right)
\frac{e^{-t\, H_C}}{\tilde{m}}\, ,
\label{intform}
\end{eqnarray}
where $\gamma_v = 1/\sqrt{1-v^2/c^2}$,
then relation (\ref{35abb}) becomes
%
%
%
\begin{widetext}
\begin{eqnarray}
0\ &=& \ \int_{0}^{\infty}\!\!\rmd \tilde{m} \
\! f_{\frac{1}{2}}\!\left(\tilde{m},  t\,c^2,  t\,c^2m^2 \right)
\, \langle {\bi x}''| e^{-(t''- \tau)\,\hat{H}_C}
\left(\frac{\delta_{ij}}{\tilde{m}} + [\dot{\hat{\bi x}}_i, \hat{\bi x}_j]\right)e^{-(\tau- t')\,\hat{H}_C}
|{\bi x}' \rangle \nonumber \\[2mm]
&=& \  \langle {\bi x}''| e^{-(t''- \tau)\hat{H}_{SR}}\left(\frac{\delta_{ij}}
{m \gamma_v} + [\dot{\hat{\bi x}}_i, \hat{\bi x}_j]\right)e^{-(\tau- t')\hat{H}_{SR}}
|{\bi x}'\rangle  \, .
\label{35ab}
\end{eqnarray}
\end{widetext}
Similarly for negative frequencies we obtain [see \cite{JK2}, Section~3]
\begin{eqnarray}
\mbox{\hspace{-3mm}}&&\langle {\bi x}''| e^{-(t''- \tau)\hat{H}_{SR}}
\left(\frac{\delta_{ij}}{m \gamma_v} - [\dot{\hat{\bi x}}_i, \hat{\bi x}_j]\right)e^{-(\tau- t')
\hat{H}_{SR}}|{\bi x}'\rangle\nonumber \\
\mbox{\hspace{-3mm}}&&\ = \ 0\, .
\label{36abc}
\end{eqnarray}
Eq.~(\ref{35ab}) together with (\ref{36abc}) implies the matrix-valued
commutator $[\dot{\hat{\bi x}}_i, \hat{\bi x}_j] = - \sigma_3\delta_{ij}/(m \gamma_v)$.
Because of $\dot{\hat{\bi x}}_i = \sigma_3\partial \hat{H}_{SR}({\bi p})/\partial {\bi p}_i =
\sigma_3\ \!\hat{\!\bi p}_i/(m \gamma_v)$, which means
$\hat{\bi p}_i = \sigma_3 m \gamma_v \dot{\hat{\bi x}}_i$,
we can make use of the algebraic identity
\begin{eqnarray}
[\hat{\bi x}_j, \hat{\bi p}_i] \ = \ \sigma_3 m \gamma_v [\hat{\bi x}_j,\dot{\hat{\bi x}}_i] \ +
\ \sigma_3 m \! \ \dot{\hat{\bi x}}_i[\hat{\bi x}_j, \gamma_v]
\, .
\end{eqnarray}
Reminding now that
\begin{eqnarray}
[\hat{\bi x}_j, \gamma_v(\dot{\hat{\bi x}}_i)] =
[\hat{\bi x}_j, \dot{\hat{\bi x}}_i]\cdot\frac{\partial \gamma_v}{\partial \dot{\hat{\bi x}}_i}\, ,
\end{eqnarray}
we finally arrive at
\begin{eqnarray}
[\hat{\bi x}_j, \hat{\bi p}_i] \ = \ \delta_{ij} + \frac{\hat{\bi p}_i \hat{\bi p}_j}{m^2 c^2}\, .
\end{eqnarray}
The analytical continuation of the relativistic stochastic process (\ref{8a})
gives the corresponding Quantum Mechanical CCR
\begin{eqnarray}
[\hat{\bi x}_j, \hat{\bi p}_i]_{\rm SR} \ = \
\rmi\left(\delta_{ij} + \frac{\hat{\bi p}_i \hat{\bi p}_j}{m^2 c^2}\right)\, .
\label{36abb}
\end{eqnarray}
This coincides with the SR commutator that one obtains by lifting Dirac brackets
(corresponding to the first class constraint $\Phi \equiv p^2 - m^2$ and the gauge
condition $\chi \equiv x^{\mu}p_{\mu} - \varsigma m^2c^2$, with $\varsigma$ being
the world-line parameter) to QM commutators. The gauge condition leading to (\ref{36abb})
is precisely the Polyakov gauge condition which, as mentioned earlier, is implicit
in our  superstatistics formulation~\cite{JK2,note}. It is worth of noting that the
SR commutator (\ref{36abb}) appears frequently when the first quantization of relativistic
systems is discussed, see e.g., Refs.~\cite{Bette:80,Ghosh:07,Ghosh:06,Banerjee:06}.

The roughness of a typical relativistic path can be evaluated by rewriting (\ref{35ab})
in a time-sliced version as
\begin{eqnarray}
&&\mbox{\hspace{-7mm}}\frac{\Delta t}{m}\ \!\langle {\bi x}'',t''|\! \
\sqrt{1-\frac{(\hat{\bi x}_i(t + \Delta t) - \hat{\bi x}_i(t))^{2}}{(\Delta t)^2 c^2}}
\ |{\bi x}',t'\rangle \nonumber \\[1mm]
&&\mbox{\hspace{-5mm}}\;\;\;\;\;\;\;\; = \ \langle {\bi x}'',t''|\! \ (\hat{\bi x}_i(\tau +
\Delta t) - \hat{\bi x}_i(\tau))^{2}\! \ |{\bi x}',t'\rangle \, ,
\end{eqnarray}
which gives for $\Delta \hat{\bi x}_i(\tau) \equiv \hat{\bi x}_i(\tau + \Delta t) - \hat{\bi x}_i(\tau)$
\begin{eqnarray*}
c \Delta t  =    \langle {\bi x}'',t''|\! \
|\Delta \hat{\bi x}_i(\tau)|\sqrt{1+m^2c^2(\Delta \hat{\bi x}_i(\tau))^2}  \ |{\bi x}',t'\rangle\, .
\end{eqnarray*}

In particular, for $\Delta x$ much smaller than the Compton wavelength $\lambda_C = 1/mc$,
the Hausdorff dimension $d_H =1$, i.e. for such short times the process is super-diffusive.
In the opposite case, when $\Delta x$ is much bigger
than $\lambda_C$ we recover the non-relativistic Feynman--Hibbs scaling (\ref{B.34})
with $d_H = 2$.

%
%
\section*{Appendix C~\label{apc}}

%
In this Appendix we prove the relations (\ref{31a}) and (\ref{31aaa}).
In both cases we start from relation (\ref{22a}), which we take for granted.\\

{\em a) Relation (\ref{31a}).} --- The DSR1 Hamiltonian (\ref{HDSR1}) can be
brought into the same form as the relativistic Hamiltonian $c\sqrt{{\bi p}^2 + m^2c^2}~$
by renaming the speed of light
\begin{eqnarray}
\bar{c}^2 \equiv c^2 \lambda\, ,
\end{eqnarray}
with
\begin{eqnarray}
\lambda = \frac{1}{1-m^2c^2\ell^2}\, .
\end{eqnarray}
Then
\begin{eqnarray}
H_{DSR1} \ &=& \ c\ \!\frac{{- m^2 c^2 \ell  + \sqrt{{\bi p}^2(1-m^2 c^2\ell^2) +
m^2 c^2}}}{1-m^2c^2\ell^2}\nonumber\\[2mm]
&=& - m^2 c \ell \bar{c}^2 + \bar{c}\sqrt{{\bi p}^2 + m^2 \bar{c}^2}\, ,
\label{Hbar1}
\end{eqnarray}
where in the last line $c$ is understood to be expressed as a function of $\bar{c}$,
precisely $c=\bar{c}/\sqrt{1 + m^2\ell^2\bar{c}^2}$.
Now, relation (\ref{22a}) must hold also when we formally replace everywhere
$c$ with $\bar{c}$. Therefore
\begin{widetext}
\begin{eqnarray}
 &&\mbox{\hspace{-15mm}}\int_{\footnotesize{\bi x}(0)\ \! =
 \ \! \footnotesize{\bi x}'}^{\footnotesize{\bi
x}(t) \ \! = \ \! \footnotesize{\bi x}} \ \!{\mathcal{D}}{\bi x} \frac{\mathcal{D}
{\bi p}}{(2\pi)^D} \ \! \exp\left\{\int_{0}^{t} \!\!\rmd \tau \
\!\left[\rmi {\bi p}\cdot \dot{\bi x} \ - \
\bar{c}\sqrt{{\bi p}^2 + m^2 \bar{c}^2}\right]\right\} \nonumber \\[2mm]
&&\mbox{\hspace{15mm}}= \  \int_{0}^{\infty}\!\!\rmd \tilde{m} \
\! f_{\frac{1}{2}}\!\left(\tilde{m}, tc^2\lambda, tc^2\lambda m^2\right)
\int_{\footnotesize{\bi x}(0)\ \! = \ \! \footnotesize{\bi x}'}^{\footnotesize{\bi
x}(t) \ \! = \ \! \footnotesize{\bi x}} \ \!{\mathcal{D}}{\bi x} \ \! \frac{\mathcal{D} {\bi
p}}{(2\pi)^D} \ \! \exp\left\{\int_{0}^{t} \!\!\rmd \tau\
\!\left[\rmi {\bi p}\cdot \dot{\bi x} \ - \  \frac{{\bi p}^2}{2
\tilde{m}} \ - \  m \bar{c}^2 \right]\right\}\, .
\label{23aa}
\end{eqnarray}
\end{widetext}
But
\begin{eqnarray}
m\bar{c}^2 \ = \ \frac{mc^2}{1-m^2c^2\ell^2} \ = \ m^2 c\ell\bar{c}^2 + \frac{mc^2}{1+mc\ell}\, ,
\label{mc2}
\end{eqnarray}
therefore, defining $E_0=mc^2/(1+mc\ell)$, the relation (\ref{23aa}) becomes
\begin{widetext}
\begin{eqnarray}
 &&\mbox{\hspace{-15mm}}\int_{\footnotesize{\bi x}(0)\ \! =
 \ \! \footnotesize{\bi x}'}^{\footnotesize{\bi
x}(t) \ \! = \ \! \footnotesize{\bi x}} \ \!{\mathcal{D}}{\bi x} \frac{\mathcal{D}
{\bi p}}{(2\pi)^D} \ \! \exp\left\{\int_{0}^{t} \!\!\rmd \tau \
\!\left[\rmi {\bi p}\cdot \dot{\bi x} \ + m^2 c\ell\bar{c}^2 \ - \
\bar{c}\sqrt{{\bi p}^2 + m^2 \bar{c}^2}\right]\right\} \nonumber \\[2mm]
&&\mbox{\hspace{15mm}}= \  \int_{0}^{\infty}\!\!\rmd \tilde{m} \
\! f_{\frac{1}{2}}\!\left(\tilde{m}, tc^2\lambda, tc^2\lambda m^2\right)
\int_{\footnotesize{\bi x}(0)\ \! = \ \! \footnotesize{\bi x}'}^{\footnotesize{\bi
x}(t) \ \! = \ \! \footnotesize{\bi x}} \ \!{\mathcal{D}}{\bi x} \ \! \frac{\mathcal{D} {\bi
p}}{(2\pi)^D} \ \! \exp\left\{\int_{0}^{t} \!\!\rmd \tau\
\!\left[\rmi {\bi p}\cdot \dot{\bi x} \ - \  \frac{{\bi p}^2}{2
\tilde{m}} \ - \ E_0 \right]\right\}\, ,
\end{eqnarray}
\end{widetext}
which, in force of relation (\ref{Hbar1}), coincides with Eq.~(\ref{31a}).
Note that we can obtain the DSR expressions for $\langle \tilde{m} \rangle$ and for var$(\tilde{m})$
reported in the text, by simply replacing $c^2 \to \bar{c}^2\equiv c^2\lambda$ into the similar
expressions given for the special-relativistic case after formula (\ref{22a}).

As for the fractal dimension of the representative trajectories (or histories), one finds that
(following the same reasonings as in Appendix~B), the following scaling holds
\begin{eqnarray*}
\mbox{\hspace{-4mm}}&&c \Delta t = \langle {\bi x}'',t''|\! \
|\Delta \hat{\bi x}_i(\tau)|\sqrt{\frac{1+\lambda c^2m^2(\Delta \hat{\bi x}_i(\tau))^2}{{\lambda}}}
\ |{\bi x}',t'\rangle\, ,
\end{eqnarray*}
and thus for $\Delta x \ll \lambda_C/\sqrt{\lambda}$ we have $d_H = 1$, while for
$\Delta x \gg \lambda_C/\sqrt{\lambda}$ holds $d_H = 2$. Hence the representative
trajectories have a shorter typical length scale $\Delta x$
(i.e., average distance between change of direction) than in the SR case.
\\

{\em b) Relation (\ref{31aaa}).} --- We can prove relation (\ref{31aaa}) in an analogous way, by noting
that the DSR2 Hamiltonian (\ref{28bba}) can be written as
\begin{eqnarray}
H_{DSR2} \ = \ \! \frac{\sqrt{{\bi p}^2c^2 + m^2 c^4}}{\sqrt{1 + m^2 c^2 \ell^2}} \ = \
\sqrt{{\bi p}^2\bar{c}^2 + \bar{m}^2\bar{c}^4}\, ,
\label{Hbar2}
\end{eqnarray}
once we make the replacements
\begin{eqnarray}
&&\mbox{\hspace{-8mm}}c^2 \to \,\,\,\bar{c}^2  \ \equiv \ \frac{c^2}{1+m^2c^2\ell^2} \ =
\ c^2\zeta^2\, , \nonumber\\[4mm]
&&\mbox{\hspace{-8mm}}m^2 \to \bar{m}^2  \ \equiv \  m^2(1+m^2c^2\ell^2) \ = \ m^2/\zeta^2\, ,
\label{rep}
\end{eqnarray}
where $\zeta = 1/\sqrt{1 + m^2 c^2 \ell^2}$.
Therefore relation (\ref{22a}) must formally hold once we make such replacements in it,
and these bring to the following expressions
\begin{eqnarray}
&&\bar{m}\bar{c}^2 \ = \ \frac{mc^2}{\sqrt{1 + m^2 c^2 \ell^2}} \ = \ \bar{E}_0\, , \nonumber\\[4mm]
&&f_{\frac{1}{2}}\!\left(\tilde{m}, t\bar{c}^2, t\bar{c}^2\bar{m}^2\right) \ = \
f_{\frac{1}{2}}\!\left(\tilde{m}, tc^2\zeta^2, tc^2m^2\right)\, , \nonumber\\[4mm]
&&\langle\tilde{m}\rangle  \ = \  \bar{m} + \frac{1}{t \bar{c}^2} \ = \
\frac{m}{\zeta}  +  \frac{1}{tc^2\zeta^2}\, ,\nonumber\\[4mm]
&&{\rm var}(\tilde{m}) \ = \ \frac{\bar{m}}{t\bar{c}^2} + \frac{2}{t^2 \bar{c}^4} \ = \
\frac{m}{tc^2\zeta^3} + \frac{2}{t^2c^4\zeta^4}\, ,
\end{eqnarray}
which prove the relation (\ref{31aaa}) and the expressions for
$\langle\tilde{m}\rangle$ and var$(\tilde{m})$ as given in the main text.
$\bar{E}_0$ is identified with the particle's rest energy.

As for the scaling behavior $\Delta x$ vs. $\Delta t$ we find in this case
\begin{eqnarray*}
\mbox{\hspace{-4mm}}&&c \zeta \Delta t = \langle {\bi x}'',t''|\!
\ |\Delta \hat{\bi x}_i(\tau)|\sqrt{{1+c^2m^2(\Delta \hat{\bi x}_i(\tau))^2}}  \ |{\bi x}',t'\rangle\, ,
\end{eqnarray*}
which provides the same fractal dimension as SR, i.e.,
for $\Delta x \ll \lambda_C$ one has $d_H = 1$, while for $\Delta x \gg \lambda_C$ holds $d_H = 2$.
%
%

\section*{Appendix D~\label{apd}}

%
Similar arguments as in Appendix B can be now applied to the DSR systems
(\ref{HDSR1}) and (\ref{28bba}) to compute the fundamental commutators.
In the DSR system~(\ref{HDSR1}) we consider the replacement
\begin{eqnarray}
c^2 \rightarrow \bar{c}^2\ \equiv \ c^2\lambda\, ,
\end{eqnarray}
and because of relations (\ref{Hbar1}) and (\ref{mc2}) of Appendix C,
we can rewrite the identity (\ref{49}) as
\begin{eqnarray}
\mbox{\hspace{-2mm}}e^{-t\, H_{DSR1}} \ = \
\int_{0}^{\infty}\rmd \tilde{m} \
\! f_{\frac{1}{2}}\!\left(\tilde{m},  t\, \bar{c}^2,  t\, m^2\bar{c}^2\right)
e^{- t\, H_{C1}}\, ,
\end{eqnarray}
with
\begin{eqnarray}
H_{C1}  \ = \ \frac{{\bi p}^2}{2\tilde{m}} \ +
\  E_0 \quad {\rm and} \quad E_0 \ = \ \frac{mc^2}{1+mc\ell}\,.
\end{eqnarray}
\\
\noindent In the DSR1 framework the Ward identity (\ref{32aa}) reads
\begin{eqnarray}
\!\!\!\!\!\!\!\!\!\!&&\int_{0}^{\infty}\!\!\!\rmd \tilde{m} \
\! f_{\frac{1}{2}}\!\left(\tilde{m},t\, \bar{c}^2,  t\, m^2\bar{c}^2 \right)
\langle {\bi x}'',t''| \frac{\delta_{ij}}{\tilde{m}} +
[\dot{\hat{\bi x}}_i, \hat{\bi x}_j]|_\tau |{\bi x}',t'\rangle \nonumber \\
\!\!\!\!\!\!\!\!\!\!&&\,\,\, = \ 0\, ,
\label{62}
\end{eqnarray}
where the vectors $|{\bi x}',t'\rangle$ and the commutator $[\dot{\hat{\bi x}}_i, \hat{\bi x}_j]|_\tau$
evolve analogously as in (\ref{evol}), but now controlled by the Hamiltonian $\hat{H}_{C1}$.
The integration formula (\ref{intform}), after the replacement $c^2 \to \bar{c}^2$, reads
\begin{eqnarray}
\frac{e^{- t\, H_{DSR1}}}{m\,\gamma_v(\bar{c})} = \int_{0}^{\infty}\!\!\!\!\!\rmd \tilde{m}
\ f_{\frac{1}{2}}\!\left(\tilde{m}, t\, \bar{c}^2, t\, \bar{c}^2m^2\right)
\frac{e^{- t\, H_{C1}}}{\tilde{m}}\, ,
\end{eqnarray}
where $\gamma_v(\bar{c}) = 1/\sqrt{1 - {\bi v}^2/(c^2 \lambda)}$.

The DSR1 Ward identity (\ref{62}) then becomes
\begin{widetext}
\begin{eqnarray}
0 \ &=& \  \int_{0}^{\infty}\!\!\rmd \tilde{m} \
\! f_{\frac{1}{2}}\!\left(\tilde{m},  t\, c^2\lambda, t\, m^2c^2\lambda \right)
\, \langle {\bi x}''| e^{-(t''- \tau)\hat{H}_{C1}}
\left(\frac{\delta_{ij}}{\tilde{m}} + [\dot{\hat{\bi x}}_i, \hat{\bi x}_j]\right)e^{-(\tau- t')
\hat{H}_{C1}}|{\bi x}'\rangle \ \nonumber \\[2mm]
&=&\  \langle {\bi x}''| e^{-(t''- \tau)\hat{H}_{DSR1}}
\left(\frac{\delta_{ij}}{m \gamma_v(\bar{c})} + [\dot{\hat{\bi x}}_i, \hat{\bi x}_j]
\right)e^{-(\tau- t')\hat{H}_{DSR1}}|{\bi x}'\rangle \, .
\end{eqnarray}
\end{widetext}

Because $\dot{\hat{\bi x}}_i = \partial \hat{H}_{DSR1}({\bi p})/\partial {\bi p}_i =
\hat{\!\bi p}_i/(m \gamma_v(\bar{c}))$,
which means $\hat{\!\bi p}_i = m \gamma_v(\bar{c})\dot{\hat{\bi x}}_i$,
we see that the corresponding CCR can be
written (after analytical continuation) in the form
\begin{eqnarray}
[\hat{\bi x}_j, \hat{\bi p}_i]_{\rm DSR1} \ &=&
\ \rmi\left(\delta_{ij} + \frac{\hat{\bi p}_i \hat{\bi p}_j}{m^2 c^2\lambda}\right) \nonumber \\[2mm]
&=& \ [\hat{\bi x}_j, \hat{\bi p}_i]_{\rm SR} \ - \ \rmi \ \! {\ell}^2 \hat{\bi p}_i \hat{\bi p}_j \, .
\label{39abb}
\end{eqnarray}
In the low momentum limit (i.e. when $|{\bi p}|\ll 1/\ell$) the commutator
$[\hat{\bi x}_j, \hat{\bi p}_i]_{\rm DSR1}$ approaches the  SR commutator (\ref{36abb}).
In passing we note that the CCR  (\ref{39abb}) resembles the
Snyder commutators (\ref{Snyder}) [see also Ref.~\cite{Snyder:47}], which are familiar commutators
of DSR~\cite{Ghosh:07,Ghosh:06,Banerjee:06}.

In the second case (cf. Eq.~(\ref{28bba})) we use the replacements (\ref{rep}) [see again Appendix C]
so that identity (\ref{49}) now, in the DSR2 framework, becomes
\begin{eqnarray}
&&\mbox{\hspace{-10mm}}e^{-t\, H_{DSR2}}  =  \int_{0}^{\infty}\!\!\rmd \tilde{m} \
\! f_{\frac{1}{2}}\!\left(\tilde{m},  t\, c^2\zeta^2,  t\, m^2c^2\right)
e^{- t\, H_{C2}}\, ,
\end{eqnarray}
with
\begin{eqnarray}
\mbox{\hspace{-5mm}}H_{C2} \ = \ \frac{{\bi p}^2}{2\tilde{m}} \ + \  \bar{E}_0 \quad {\rm and} \quad
\bar{E}_0 \ = \ \frac{mc^2}{\sqrt{1+m^2c^2\ell^2}}\, .
\end{eqnarray}
%
After the replacements $c^2 \to \tilde{c}^2=c^2\zeta^2$,
$m^2 \to \bar{m}^2=m^2/\zeta^2$, and the corresponding changes in formula (\ref{intform}),
the DSR2 Ward identity finally becomes
%
\begin{eqnarray}
\!\!\!\!\!\!\!\!&&\langle {\bi x}''| e^{-(t''- \tau)\hat{H}_{DSR2}}
\left(\frac{\zeta\,\delta_{ij}}{m \gamma_v(\tilde{c})} + [\dot{\hat{\bi x}}_i, \hat{\bi x}_j]\right)
e^{-(\tau- t')\hat{H}_{DSR2}}|{\bi x}'\rangle \nonumber \\
\!\!\!\!\!\!\!\!&&= \ 0 \, .
\end{eqnarray}
where $\gamma_v(\tilde{c}) = 1/\sqrt{1 - {\bi v}^2/\tilde{c}^2}$.

Since $\dot{\hat{\bi x}}_i = \partial \hat{H}_{DSR2}({\bi p})/\partial {\bi p}_i =
\hat{\!\bi p}_i/(\bar{m} \gamma_v(\tilde{c}))$,
we see that the resulting QM CCR read
(after analytical continuation)
\begin{eqnarray}
\!\!\!\!\!\![\hat{\bi x}_j, \hat{\bi p}_i]_{\rm DSR2} \ =
\ \rmi\left(\delta_{ij} + \frac{\hat{\bi p}_i \hat{\bi p}_j}{m^2 c^2}\right) \, .
\label{41ba}
\end{eqnarray}
%

Note, in particular, that the CCR (\ref{41ba}) coincides with the SR commutator (\ref{36abb}).
This fact should not be so
surprising, since CCR's directly reflect the roughness of the representative paths~\cite{note2}
and from Appendix~C we know that the fractal dimension of the DSR2 system coincides with that of SR.
%
%

%
%
\section*{References}

%

\begin{thebibliography}{99}
\bibitem{note} In general, the canonical gauge $\chi(x,p)$ must in SR explicitly depend on the world-line
parameter $\varsigma$. This is because the constraints must be preserved under evolution and thus
particularly $0\equiv \rmd\chi/\rmd\varsigma =\partial{\chi}/\partial{\varsigma} +\frac{1}{2} \{\chi, e \Phi\} =
\partial{\chi}/\partial{\varsigma} + e p^{\mu}\partial{\chi}/\partial{x^{\mu}}$ ($e$ is the einbein).
Our choice of the gauge is non-canonical and 
coincides with the Polyakov or temporal gauge~\cite{polyakov:87}: $\chi = \rmd e/ \rmd\varsigma$.
\bibitem{note2} We have already seen that the Feynman--Hibbs scaling relation (from which follows the
Hausdorff dimension of representative histories) results from Ward identity. At the same time the
CCR directly follows from the Feynman--Hibbs scaling relation~\cite{Kroger}.
\end{thebibliography}
\end{document}